\documentclass[12pt,preprint]{aastex}

\begin{document}

\date{\today}

\title{The cluster ages experiment (CASE). V. Analysis of three eclipsing binaries 
       in the globular cluster M4{\LARGE$^\ast$}
      }

\author{J. Kaluzny$^1$, I. B. Thompson$^2$, M. Rozyczka$^1$, 
        A. Dotter$^3$, W. Krzeminski$^1$, W. Pych$^1$, S. M. Rucinski$^4$, 
        G. S. Burley$^2$, and S. A. Shectman$^2$}
\affil{$^1$\small{Nicolaus Copernicus Astronomical Center, Bartycka 18, 00-716 Warsaw, 
      Poland}; \footnotesize{jka@camk.edu.pl, mnr@camk.edu.pl, wk@camk.edu.pl,
      batka@camk.edu.pl,pych@camk.edu.pl}}
\affil{$^2$\small{Observatories of the Carnegie Institution of Washington, 813 Santa 
       Barbara Street, Pasadena, CA 91101-1292, USA}; 
       \footnotesize{ian@obs.carnegiescience.edu, burley@obs.carnegiescience.edu,
       shec@obs.carnegiescience.edu}}
\affil{$^3$\small{Space Telescope Science Institute, 3700 San Martin Drive, 
       Baltimore, MD 21218, USA}; \footnotesize{aaron.dotter@gmail.com}}
\affil{$^4$\small{Department of Astronomy \& Astrophysics, David Dunlap Observatory, 
       University of Toronto PO Box 360, Richmond Hill, ON L4C 4Y6, Canada}; 
       \footnotesize{rucinski@astro.utoronto.ca}}
%\email{}

\begin{abstract}
We use photometric and spectroscopic observations of the eclipsing binaries V65, V66 and V69 in the 
field of the globular cluster M4 to derive masses, radii, and luminosities of their components. 
The orbital periods of these systems are 2.29, 8.11 and 48.19 d, respectively. The 
measured masses of the primary and secondary components ($M_p$ and $M_s$)   are  
0.8035$\pm$0.0086 and 0.6050$\pm$0.0044 M$_\odot$ for V65,  0.7842$\pm$0.0045 and 
0.7443$\pm$0.0042 M$_\odot$ for V66, 
and 0.7665$\pm$0.0053 and 0.7278$\pm$0/0048 M$_\odot$ for V69. The measured radii 
($R_p$ and $R_s$) are 1.147$\pm$0.010 and 0.6110$\pm$0.0092 R$_\odot$ for V66, 
0.9347$\pm$0.0048 and  0.8298$\pm$0.0053 R$_\odot$ for V66, and 0.8655$\pm$0.0097 and 
0.8074$\pm$0.0080 R$_\odot$ for V69.
The orbits of V65 and V66 are circular, whereas that of V69 has 
an eccentricity of 0.38. Based 
on systemic velocities and relative proper motions, we show that all the three systems are members 
of the  cluster. We find that the distance to M4 is 1.82$\pm$0.04 kpc - in good 
agreement with recent estimates based on entirely different met\-hods. 
We compare the absolute parameters of V66 and V69 with two sets of theoretical isochrones in mass-radius
and mass-luminosity diagrams, and for an assumed [Fe/H] = -1.20, [$\alpha$/Fe] = 0.4, and Y = 0.25 we
find the most probable age of M4 to be between 11.2 
and 11.3 Gyr. CMD-fitting with the same parameters yields an age close to, or slightly in excess of,
12 Gyr. However, considering the sources of uncertainty involved in CMD fitting, these two methods of 
age determination are not discrepant. Age and distance determinations can be further improved when 
infrared eclipse photometry is obtained.


\end{abstract}


\section{Introduction}
\label{sect:intro}
Detached eclipsing double-line binaries (DEBs) are the primary source of the observational 
data concerning stellar masses and radii. When supplemented by luminosities derived from
parallaxes, empirical relations between color and effective temperature, or fits to 
disentangled spectra, they enable fundamental tests of stellar evolution models. 
For many 
field Population I binaries with components at solar mass or larger, modern high accuracy 
measurements of masses, radii and luminosities are in general agreement with theoretical 
predictions \citep[see, for example,][]{lac05,lac08,cla08}. Similar encouraging results 
are obtained for binaries in the old open clusters NGC 188 \citep{mei09}, NGC 2243 
\citep{jka06}, and NGC 6791 \citep{gru08,bro11}. 
On the other hand, the models 
seem to underestimate the radii of numerous K and M dwarfs in short-period binaries.
Summaries of relevant recent 
measurements can be found for example in \citet{bla08}, \citet{tor10b}, and \citet{kra11}.
\let\thefootnote\relax\footnotetext
{$^{\mathrm{\ast}}$This paper includes data gathered with the 6.5-m Magellan Baade 
and Clay Telescopes, and the 2-5 m du Pont Telescope located at Las Campanas Observatory, 
Chile.}

The situation is less clear for Population II stars, for which only a very few DEBs 
with main-sequence components are known \citep{tho10}, and there is an urgent need to 
locate and study such systems. Within the series Cluster AgeS Experiment (CASE), this is 
the second paper devoted to the study of globular-cluster DEBs with main-sequence or 
subgiant components. The general goal of CASE is to determine the basic stellar parameters 
(masses, luminosities, and radii) of the components of cluster binaries to a precision 
better than 1\% in order to measure cluster ages and distances, and to test stellar 
evolution models \citep{jka05}. The methods and assumptions we employ utilize basic and 
simple approaches, following the ideas of \citet{pac97} and \citet{tho01}. Previous 
CASE papers analyzed blue straggler systems in $\omega$ Cen \citep{jka071} and 47 Tuc 
\citep{jka072}, an SB1 binary in NGC 6397 \citep{jka08} and the binary V69-47~Tuc, which is 
an SB2 system with main-sequence components \citep{tho10}.

The present paper is devoted to the analysis of three DEBs, V65, V66 and V69, 
all members of the globular cluster M4. We use  radial velocity and photometric 
observations to determine accurate masses, luminosities, and radii of 
the components of these systems.

Section \ref{sect:photobs} 
describes the photometric observations and the determination of orbital ephe\-merides. Section 
\ref{sect:spectobs} presents the spectroscopic observations and the radial-velocity 
measurements. The combined photometric and spectroscopic solutions for orbital elements and 
component parameters are obtained in Section \ref{sect:lca_sp}, while the cluster membership 
of the three DEBs is discussed in Section \ref{sect:member}. In Section \ref{sect:age}, we 
compare the derived parameters to a selection of stellar evolution models, with an emphasis 
on estimating the age of the system. Finally, in Section \ref{sect:discus} we summarize our 
findings.

\section{Photometric observations}
\label{sect:photobs}

Our survey for eclipsing binaries in M4 began in July 1995 with a 2-week observational 
campaign at CTIO. We monitored the cluster in the $B$- and $V$-bands, using the 2K$^2$ TEK2 
camera attached to the 0.9-m telescope. A single eclipse of V66 was detected, occurring 
on 1995 July 18 UT. 
The survey continued on the 1.0-m Swope telescope at Las Campanas Observatory 
(LCO) using five different CCD 
cameras (Ford, TEK3, SITe1, SITe2, and SITe3) over several seasons during the period  1996 - 2009. 
Most of the observations were obtained with the SITe3 camera, with a plate scale of 
0.435 arcsec/pixel and a field size of 14.8$\times$22.8 arcmin$^2$. Some early results from
this survey were presented in  \citet{jka97}.

The first eclipses of V65 and V69 were
detected on 1996 April 16 UT and 1996 April 21 UT, respectively. In the same year preliminary but 
reliable ephemerides for V65 and V66 were determined.
A part of a secondary eclipse of V69 was observed on 1998 August 17 UT. However, because
of the relatively long period for this system, the data 
were still insufficient to establish a unique ephemeris. An initial ephemeris was derived from 
constraints provided by substantial out-of-eclipse photometry performed in 1996 - 2005, and 
radial velocity observations obtained with the MIKE spectrograph on the Magellan Baade and
Clay telescopes  (see Section \ref{sect:spectobs}). 
From 1998 until 2009 we also observed our targets with the 2.5-m du Pont telescope 
at LCO equipped with the TEK5 2K$^2$ camera with a plate scale of 0.259 arcsec/pixel. Between 2001 
and 2009, several eclipses were covered for V65 and V69, and some out-of-eclipse data were 
collected for V69. In all observing runs, on each telescope, the same $B$ and $V$ filters 
were used. Times of minima are presented in Tables \ref{v10989_min}, \ref{v11888_min},
and \ref{v23900_min}.

Linear ephemerides provide adequate 
fits to the photometric data for all three binaries, these are given by Equations (1). In the case 
of V65 and V66 the ephemerides were obtained from the moments of minima calculated from individual 
light curves. This procedure could not be applied to V69, as in only three eclipses both the 
descending and the ascending branch was observed. In the case of V69, the period was 
derived using the algorithm developed by \citet{laf65}, and the moment of the primary eclipse 
was found from the phased light curve with an improved version of the method developed by 
\citet{kwee}. For V65, only  du Pont observations were used,  reduced with the image 
subtraction technique. For V66, we also made use of photometry measured
with  profile fitting in images collected with the Swope and 0.9-m CTIO telescopes.
\begin{eqnarray}
 HJD^\mathrm{V65}_\mathrm{min}& = &244 9905.49577(43) + 2.29304564(26)\times E\nonumber\\
 HJD^\mathrm{V66}_\mathrm{min}& = &244 9900.41965(26) + 8.11130346(85)\times E
 \label{eq:ephem}\\
 HJD^\mathrm{V69}_\mathrm{min}& = &245 0048.34890(14) + 48.1882687(6)\times E\nonumber
\end{eqnarray}

Table \ref{eqcoord} gives the equatorial coordinates of the three variables analysed in this paper. 
They are tied to the UCAC3 system \citep{zach10} for V65 and V66 , and  to 
the GSC-1.0 system \citep[e.g.][]{las90} for V69. Finding charts prepared from du Pont TEK5 
$V$-band images are shown in Figure \ref{fig:charts}.

V65 is blended with two stars located at angular distances of 0.37\arcsec\ and 0.57\arcsec. 
The variable is the brightest component of the blend. All three stars are included in ACS-HST 
HST F606W/F814W photometry published by \citet{and08}, with star identification
numbers 13362, 13372 (V65), and 13374. The HST photometry is listed in Table \ref{and_phot}.
We determined the rectangular coordinates of V65 and its two close visual 
companions on the reference image by transforming their positions from ACS/HST 
photometry of \citet{and08}, and the presence of the companions was taken into account while
extracting light curves of this binary. V66 is listed by \citet{and08} as star 5795. 
It does not suffer from any blending which could affect the ground based photometry. V69 
is not present in any of the currently available HST images of M4. The object is located at an 
angular distance of 7.69\arcmin\ from the center of the cluster, beyond the half-light 
radius of 4.33\arcmin\ \citep[][2010 edition]{har96}. The  du Pont
images show no evidence of any unresolved visual companions to V69.

\subsection{Light curves and calibration of photometry.}
\label{sect:lccal}

The $B$ and $V$ light curves of V65 and V66 were determined entirely from the 
du Pont data. They were extracted with the image subtraction technique, using methods 
and codes described
in \citet{jka10}. In short, we used a combination of ISIS 2.1 \citep{alard98}, 
Allstar/Daophot \citep{ste87} and Daogrow \citep{ste90} codes, supplemented with some 
IRAF$^1$\footnote {$^1$IRAF is distributed by the National Optical Astronomy Observatories,
which are operated by the AURA, Inc., under cooperative agreement  with the NSF.} 
tasks. For these two variables, we also extracted light curves using  
profile fitting software to make sure that the image subtraction technique is free from 
systematic errors (in particular, we checked that the curves resulting from profile
photometry and image subtraction have the same amplitudes). As expected, the scatter 
produced by the image subtraction technique was significantly smaller. 

$B$ and $V$ light curves of V69 were obtained from the data collected with the 
Swope telescope and the SITe3 camera. The light curves were measured using profile 
fitting software since
this variable is located in a sparsely populated field and  image subtraction 
offers no advantage in comparison with the traditional profile photometry. We also 
collected a few frames containing V69 with the du Pont telescope and used these 
to measure the magnitude and color of this variable at maximum light. 
Observations  \citet{land92} photometric standards obtained with the du Pont
telescope enabled us to transform the light curves of all three variables to the 
standard $BV$ system.

Full details of the procedure employed to perform the photometric measurements will be 
reported in a separate paper devoted to the photometry of other variables in M4 
(Kaluzny et al., in preparation), along with the measurement of  differential and global 
reddening in the field of the cluster.
Table \ref{BV} lists magnitudes and colors of V65, V66 and V69 at 
maximum light, together with the reddening towards each of the variables.
The errors include internal and external uncertainties, and the last 
column gives the total reddening as determined by Kaluzny et al. (in preparation;
the reddening was found by comparing turnoff colors of M4 and the low-extinction 
cluster NGC 6362).

The light curve of V65 is unstable (see Figure \ref{fig:V65all}). The fast 
rotation of the components, implied by the relatively short orbital period of 2.29~d,
apparently induces a strong magnetic activity in at least one of the stars. A clear 
sign of such an activity is the X-ray emission: the system was listed by \citet{bas04} 
as the X-ray source CX 30 with $L_{\rm x}=2.6\times10^{29}$~erg~s$^{-1}$. For the
present analysis, we selected observations 
collected between 2008 June 07 and 2009 June 30. During that period the light curve was 
flat between the eclipses, the eclipses were symmetric (note the totality
of the secondary eclipse), and the binary was brighter than in the other observing seasons. 
All these observations indicate that the magnetic activity of the system was significantly
lower than during the other observing seasons.

The remaining two light curves are stable. While the eclipses of the star 
V66 are symmetric and separated by a half of the orbital period, the orbit of  
V69 is clearly eccentric, and the secondary eclipse occurs at phase 0.609. The final light 
curves adopted for the analysis of the three systems are shown in Figure \ref{fig:3lc}. 

\section{Spectroscopic observations}
\label{sect:spectobs}

The spectra were taken with the MIKE echelle spectrograph \citep{bern03} on the 
Magellan Baade and Clay 6.5-m telescopes, using a 0.7\arcsec\ slit, which provided a resolution 
R~$\approx$40,000. A typical observation consisted of two 1800-second exposures of
the target, flanking an exposure of a thorium-argon hollow-cathode lamp. 
A few of the exposures were shorter, depending on the observing conditions.
The raw 
spectra were reduced with the pipeline software written by Dan Kelson, following 
the approach outlined in \citet{kel03}. The IRAF 
package ECHELLE was used for the post-extraction processing. 
The velocities were measured using software based on the TODCOR algorithm 
\citep{zuc94}, kindly made available by Guillermo Torres. For velocity templates, 
we used synthetic echelle-resolution spectra from the library of \citet{coelho06}.
These were interpolated to the values of $\log g$ and $T_{eff}$ derived from 
the photometric solution (see Section \ref{sect:lca_sp}) and
assuming  ${\rm [Fe/H]} = -1.2$ 
with an $\alpha$-element enhancement of 0.4 \citep{car09, dotter10}. The results of 
the velocity measurements are insensitive to minor 
changes in these parameters. 
The templates were Gaussian-smoothed to match the 
resolution of the observed spectra. In the case of V65, a rotational broadening was 
additionally applied.  

For both V66 and V69, the cross-correlations with template spectra were 
performed 
independently on wavelength intervals 4120-4320 \AA\ and 4350-4600 \AA,
covering the region of the MIKE blue spectra with the best signal-to-noise ratio
while avoiding the H~$\gamma$ line. The 
final velocities adopted for the analysis were obtained by averaging the results 
of these two measurements. In the case of V65, the spectrum was contaminated 
by one of the stars blended with the target (ID 13374 in Table \ref{and_phot}), 
so that a three-dimensional extension 
of the original TODCOR algorithm \citep{zuc95} had to be used. The cross-correlation
was performed on the interval 4000-4840~\AA\ to maximize the signal from the 
very faint secondary component. The second contaminating star (ID 13362 in Table 
\ref{and_phot}), which is fainter and more distant on the
sky from the target, was not detected in the velocity cross correlation functions.

The results of velocity measurements are presented in Tables \ref{v10989vel}, 
\ref{v11888vel}, and \ref{v23900vel}, which list heliocentric Julian dates at 
mid-exposure, velocities of the primary and secondary components, and orbital phases 
of the observations, calculated according to the ephemerides given by Equations (1). 
In the case of V65 the velocities of the contaminating star are also
given.
The observed velocity curves were fit with a nonlinear least squares solution, using 
code kindly made available by Guillermo Torres. Both the observed curves and 
the fitted ones are shown in Figure \ref{fig:3vc}, and the derived orbital parameters 
are  listed in Table \ref{orb_parm}, together with errors as returned by the fitting 
routine. Table \ref{orb_parm} also lists the velocity standard deviations 
$\sigma_p$ and $\sigma_s$
of the orbital solution which are a measure of the precision of a single
velocity measurement. In all cases, the fits adopted the periods and times of primary
eclipse as given in Equations (1).

\section{Light curve analysis and system parameters}
\label{sect:lca_sp}

The analysis of the light curves was performed with the PHOEBE implementation 
\citep{prsa05}
of the Wilson-Devinney (WD) model \citep{wil71, wil79}. The PHOEBE/WD package 
utilizes the Roche geometry to approximate the shapes of the stars, uses Kurucz model 
atmospheres, treats reflection effects in detail, and, most importantly, allows for
the simultaneous analysis of $B$ and $V$ data. The resulting geometrical parameters 
are  largely determined by the higher-quality $V$ data, while the $B$ data serves
mainly to estimate of the luminosity ratio $L_s/L_p$ in that band. To find the best 
initial parameters for PHOEBE iterations, we solved for the $V$ data using 
the JKTEBOP code \citep[][and references therein]{south04}, which, unlike PHOEBE,
can deal with a single light curve only, but it is capable of a robust search for 
the global minimum in the parameter space.

Before solving for the light curves, it was necessary to estimate the effective 
temperature of the primary, $T_p$. To that end, we used the $(B-V)$ and $E(B-V)$ 
values from Table \ref{BV}, and the calibration of \citet{cas10}. Apart from 
$T_p$, PHOEBE needs to be given a metallicity, albedo,
and gravity darkening coefficient. The user also has to specify which limb 
darkening law is to be used. We adopted a metallicity of ${\rm [Fe/H]}=-1.20$ 
(see Section  \ref{sect:spectobs} and Section \ref{sect:age}). 
Bolometric albedo and gravity darkening coefficients were set to values 
appropriate for stars with convective envelopes: $A=0.5$, and $g=0.32$ 
(we note that all three 
systems are well detached, so that effects of reflection and gravity only 
weakly affect their 
light curves). We used a linear approximation for the limb darkening. 
Theoretical limb darkening 
coefficients in the $BV$ bands were interpolated from the tables compiled by 
\citet{claret00}, using the 
jktld code.$^2$ \footnote{$^2$The code is available at 
http://www.astro.keele.ac.uk/jkt/codes/jktld.html.}
The parameters of each binary were found iteratively according to the following procedure:
\begin{enumerate}
\item Solve for the velocity curve, as explained in Section \ref{sect:spectobs}. Find a preliminary 
      solution of the $V$-light curve using JKTEBOP. Feed the obtained parameters into PHOEBE.
\item Solve for the light curves, fitting orbital inclination $i$, effective temperature of the 
      secondary $T_s$, 
      gravitational potentials at the surface of the primary $\Omega_p$ and the secondary $\Omega_s$, 
      and relative luminosities $L^B_s/L^B_p$ and $L^V_s/L^V_p$ in $B$ and $V$ bands.
\item Based on the relative luminosities obtained in Step 3, calculate the $(B-V)_0$ index of the 
      primary, and update $T_p$, using, as before, the calibrations of \citet{cas10}.
\item Repeat steps 2 and 3 until iterations converge (i.e. until the last iterated corrections to
      the parameters became smaller than the formal errors of those parameters). 
\end{enumerate}
The eccentricity and argument of periastron for V69 were found from the velocity curve. We obtained 
a preliminary photometric solution in which they were allowed to vary, but since they changed by only 
10\% of the errors given in Table \ref{orb_parm}, we decided to keep them
constant during proper iterations. 

%For all three systems, the iterations converged in two cycles. 
The residuals of the fits are shown in Figure \ref{fig:all_vars_res} and the final values of the iterated 
parameters are given in Table \ref{phot_parm}. We checked that the luminosity ratios and relative radii of 
the components were practically insensitive 
to changes in effective temperature of the primary: $(L_s/L_p)_B$, $(L_s/L_p)_V$, $r_s$, and $r_p$, 
all changed by less than 0.3\% for a $\pm$150 K change in $T_p$.
The standard errors of the six parameters iterated upon by PHOEBE were found using a 
Monte Carlo procedure written in the PHOEBE-scripter, and similar to that outlined 
in the description of the JKTEBOP code \citep[see][and references therein]{south04}. 
Briefly, the procedure replaces the observed 
light curves $B_o$ and $V_o$ with the fitted ones $B_f$ and $V_f$, generates Gaussian
perturbations $\delta B_f$ and $\delta V_f$ such that the standard deviation of the perturbation 
is equal to the standard deviation of the corresponding residuals shown in Figure~\ref{fig:all_vars_res}, 
and performs PHOEBE iterations on $B_f+\delta B_f$ and $V_f+\delta V_f$. Each Monte Carlo 
run produced 15000 points. Examples of the Monte Carlo diagrams are shown in Figure~\ref{fig:3err}.  
The combination of spectroscopic and photometric solutions yielded absolute parameters of the three DEBs 
which we list in Table \ref{abs_parm}. The 
measured masses of the primary and secondary components ($M_p$ and $M_s$)   are  
0.8035$\pm$0.0086 and 0.6050$\pm$0.0044 M$_\odot$ for V65,  0.7842$\pm$0.0045 and 
0.7443$\pm$0.0042 M$_\odot$ for V66, 
and 0.7665$\pm$0053 and 0.7278$\pm$0/0048 M$_\odot$ for V69. The measured radii 
($R_p$; $R_s$) are 1.1470$\pm$0.0104 and 0.6110$\pm$0.0092 R$_\odot$ for V66, 
0.9347$\pm$0.0048 and  0.8298$\pm$0.0053 R$_\odot$ for V66, and 0.8655$\pm$0.0097 and 
0.8074$\pm$0.0080 R$_\odot$ for V69. The measured accuracy of the mass determinations 
ranges from 0.6\% for the primary of V66 to 1.1\% for the secondary
of V65.  The measured accuracy of the radii determinations ranges from 0.5\%
for the primary of V66 to 1.5\% for the secondary of V65. The location of the 
components on the CMD of M4 is shown  in Figure \ref{fig:cmd}. 
This CMD shows data for a 3.5$\times$3.5 arcmin$^2$ field whose center is located 
1.4 arcmin from the center of the cluster. This field was chosen because 
it shows relatively uniform  extinction. The photometry of all  stars in
Figure \ref{fig:cmd} (including the 
three DEBs) has been corrected for differential reddening (Kaluzny et al., in prep.). 
No attempt was made to remove nonmembers of M4.

\section{Membership of the target binaries and the distance to M4}
\label{sect:member}
Upon averaging the subtracted images on a seasonal basis, we found no evidence
of bipolar residuals at the positions of our targets. Such residuals are
observed for objects with noticeable proper motions with respect to the
surrounding stellar field \citep{eye01}. Given the time base and the pixel scale of
our images, we rule out motions in excess of 10 mas/y. Since on the proper-motion 
diagram of M4 most of the field stars are separated from the cluster population by more than
15 mas/y \citep{zlocz12}, this is consistent with all the three targets being members~of~M4.
%0.26 arcsec/pixel; 13 years; a shift by 1/2 pixel would correspond to 
%0.13 arcsec; 0.13/13=0.01 arcsec.
 
The radial velocity of M4 is equal to 70.29$\pm$0.07 km s$^{-1}$ \citep{som09}, while the 
velocity dispersion is 3.5$\pm$0.3 km s$^{-1}$ at the core, dropping marginally towards the 
outskirts \citep{pet95}. Thus, V65 and V69 are unquestionable radial-velocity members 
of the cluster. The velocity of V66 is $\sim$2$\sigma$ larger than the cluster mean,
but the difference is too small to exclude membership. An additional argument in favor 
of cluster membership is the location of all of the components of the three DEBs 
on or very close to the main sequence  of the cluster (see Figure \ref{fig:cmd}).

Other authors have derived the distance to M4 using the Baade-Wesselink method \citep{liu90}, by 
astrometry \citep{pet95}, or by fitting the subdwarfs to the cluster's main sequence 
\citep{rich97}. Their results are remarkably consistent: 1.72$\pm$0.01, 1.72$\pm$0.14 and 
1.73$\pm$0.09 kpc, respectively \citep[see][]{rich04}; however to achieve this consistency 
it was necessary 
to replace the standard value of the ratio $R_V = A_V/E(B-V) = 3.1$ with a significantly 
larger one \citep[$R_V \sim3.8$;][]{rich04}. A thorough recent study of the reddening law 
in the field of M4 \citep{hend12} indicates that, due to the intervening Scorpius-Ophiuchus 
dark clouds, the appropriate value is $R_V=3.76\pm0.07$. Their estimate of the distance to 
the cluster, obtained from fitting of the ZAHB $V$-band magnitude to models, is 
1.80$\pm0.05$ kpc. 

Following \citet{hend12}, we adopted $R_V=3.76$, and corrected the observed magnitudes for 
extinction using the $E(B-V)$ values from Table \ref{BV}. 
To obtain absolute magnitudes in the $V$-band we used bolometric luminosities from Table 
\ref{abs_parm}, $M_{V,\odot}=4.81\pm0.03$ \citep{tor10a},
and theoretical bolometric corrections obtained from models described in Section~\ref{sect:age}. 
Having the corrected observed magnitudes and absolute magnitudes, we derived distance moduli 
separately for each component of our three DEBs. The moduli are listed in Table 
\ref{abs_parm} together with their errors which arise from uncertainties of apparent 
magnitudes from Table \ref{phot_parm}, reddening from Table \ref{BV}, $R_V$ from 
\citet{hend12}, and $M_{V,\odot}$ from \citet{tor10a}. For reasons discussed in 
Section~\ref{sect:age}, the effective temperatures of V65 components may be biased, resulting 
in luminosity errors that are hard to account for. Thus, in principle, the distance 
calculated from V66 and V69 only should be more reliable than with V65 included. With this 
restriction, we obtain a weighted mean distance (the weights being equal to inverse errors
squared) of 1.82$\pm0.04$ kpc -- in excellent agreement with the recent estimate of 
\citet{hend12}. We note that when $R_V=3.1$ is used, the distance  increases to 1.95 kpc, in disagreement 
with other estimates. Thus, our results provide an independent confirmation of the atypical 
reddening law in the field of M4.
Using all six moduli from Table \ref{abs_parm} one gets 1.85$\pm0.03$ kpc --
a value compatible with that derived from V66 and V69 alone.

\section{Isochrone age analysis} 
\label{sect:age}
The fundamental parameters derived from the binary systems reported
in this paper allow us to derive the ages of the individual stars, as
well as an aggregate age for the cluster, using theoretical isochrones.
However, before the model comparison is performed it is necessary to
review the information available on the chemical composition of M4
since the model-based ages are sensitive to the adopted values of
helium abundance, [Fe/H] and [$\alpha$/Fe].

The proximity of M4 has made it a frequent subject of spectroscopic
investigations. As summarized by \citet{iva99}, [Fe/H] determinations based
on high resolution abundance analyzes up until that time range from $-1.3$
to $-1.0$ with a mean of $-1.15$. The following is a brief review of the results
presented in large-scale spectroscopic surveys of M4:
\begin{itemize}
\item Based on the analysis of 23 stars with high resolution spectra,
\citet{iva99} derived a mean [Fe/H]=$-1.18\pm0.02$. Their measurements
 of $\alpha$-capture elements imply a mean [$\alpha$/Fe]=+0.35 (mean of O, Mg,
 Si, Ca, and Ti; note also that O and Mg exhibit significant star-to-star variations).
\item \citet{mar08} derived [Fe/H]=$-1.07\pm0.01$ and [$\alpha$/Fe]=$+0.39\pm0.05$
 from high resolution spectra of 105 stars. 
\item \citet{car09} derived [Fe/H]=$-1.200\pm0.053$ from 103 stars observed at the
 VLT with GIRAFFE (or $-1.168\pm0.066$ from 14 stars observed with UVES) where the
 quoted error is the quadrature sum of statistical and systematic errors. Further
 measurement of these same spectra yields a mean [$\alpha$/Fe]=$+0.51$ \citep[][]{car10}.
\item \citet{vil11} measured [Fe/H] = $-1.14\pm0.02$ for 23 red giant branch stars 
 located below the RGB-bump.
\item For six blue horizontal branch stars, \citet{vil12} determined [Fe/H]=$-1.06\pm0.02$.
 This value holds for the supposed subpopulation of He-enriched stars with Y=$0.29\pm0.01$.
\end{itemize}
We adopt a fiducial composition of [Fe/H]=$-1.2$, [$\alpha$/Fe]=+0.4 and Y=0.25, but will
consider a range of these parameters while deriving the ages of the binary system components.
For the following
age analysis, we use two sets of theoretical isochrones that are representative
of the state of the art for low-mass, metal-poor stars: Dartmouth \citep[][henceforth DSED]{dotter08}
and Victoria-Regina \citep[][henceforth VR]{vandenberg12}. A comparison of the derived stellar
parameters with 10, 11, 12, and 13 Gyr DSED and VR isochrones obtained for the fiducial 
composition is shown in Figure~\ref{fig:iso} in the ($M$-$R$) and ($M$-$L$) planes.
While the two sets of models are qualitatively similar, the quantitative differences 
between them lead to differences in derived ages. However, such differences are smaller
than the uncertainties imposed by the observational errors.

It is immediately clear from Figure \ref{fig:iso} that the lowest-mass star, i.e.
the secondary of V65, is larger in radius than either set of isochrones predicts, 
but that its luminosity is consistent with an (essentially) unevolved main sequence 
star. This finding is in qualitative agreement with the results discussed in 
Section~\ref{sect:intro} for nearby field binaries with similar masses, and is not 
unexpected given the dynamical and X-ray properties of V65 (see Section \ref{sect:lccal}).
The next point to notice is that the primaries of V66 and V69 yield age estimates
that are consistent with each other in both planes. The secondaries in
these systems appear to favor older ages, though not at a statistically significant
level, particularly in the ($M$-$L$) plane. This may be a consequence of the
anticorrelation of the radii in V66 and V69 illustrated in Figure \ref{fig:3err}: an
overestimate of the secondary's radius implies an underestimate of the primary's one.
If that is the case, then the age discrepancy should more-or-less cancel out when the 
average age of each system is considered.

In order to formally incorporate the observational uncertainties into the age analysis,
we evaluate the age of each star on a dense grid of points within that star's 3-$\sigma$
error box$^3$ \footnote{$^3$A 3-$\sigma$ error box represents the best compromise between fully
sampling the (assumed normal) age distribution and remaining within the parameter space
covered by the isochrone grids.} in both the mass-radius and mass-luminosity planes.
Each point at which the age is determined has a weight $w = (1+\delta)^{-1}$
where $\delta$ represents the distance of a point from the best value in ($M$-$R$) or 
($M$-$L$) plane in units of the standard deviation derived from the observations. Thus
defined, $0 < w \leq 1$. The ages and weights are used to construct weighted age 
histograms for each star.

The resulting histograms for the components of V66 and V69 are displayed in Figure~\ref{fig:hist};
V65 is omitted from the figure because its properties make it unsuitable for comparison with 
standard stellar evolution models. For completeness, the mean and standard deviations derived from 
the distributions are summarized for all stars in Table \ref{ages}. As already remarked, the DSED 
and VR isochrones give ages that agree to within one standard deviation in every case. 

Further age uncertainties are caused by sensitivity to chemical composition and inherent uncertainties 
in the stellar evolution model physics \citep[a $\sim$3\% effect, see][]{cha02}. 
We analyze the sensitivity to chemical composition in the ($M$-$R$) plane only because these quantities 
do not depend on the adopted 
composition, whereas the luminosity depends on the composition via the effective temperature. We  
calculate age differences ($\Delta$-age) with respect to the age derived assuming the fiducial 
composition. A positive $\Delta$-age value indicates that the model with varied 
composition yields an older age than the model with the fiducial composition. The numbers presented
in Table \ref{agecomp} are averaged over the components of V66 and V69; there is a slight sensitivity
of $\Delta$-age to stellar mass but it is less than 0.1 Gyr.
Increasing Y \emph{decreases} the age derived from the ($M$-$R$) plane while increasing [Fe/H]
\emph{increases} the age. Increasing the [$\alpha$/Fe] ratio also increases the derived age, but the
amount varies because of the way that $\alpha$-enhancement is defined in each set of models.
The Dartmouth models use a constant enhancement of the $\alpha$-capture elements whereas the
Victoria-Regina models employ an observationally-motivated enhancement \citep[the `GSC' heavy 
element mixture, see][]{vandenberg12}.
The age difference between scaled-solar ([$\alpha$/Fe]=0) and the $\alpha$-enhanced mixture
depends in detail on the amount to which certain elements (most notably O) are enhanced, see
\citet{vandenberg12} for a thorough discussion.

\citet{dot09} discussed the insights that may be gained by comparing stellar ages derived
from fitting the mass-radius relation of the binary V69 in 47 Tuc \citep{tho10} with those
derived from fitting isochrones to the cluster CMD. In particular, those authors showed that
the while the mass-radius diagram is sensitive to all aspects of the composition considered
above, the CMD is largely insensitive to variations in Y. Furthermore, they found that the mass-radius
and CMD ages respond differently to changes in [Fe/H]: while age and [Fe/H] are correlated in 
the ($M$-$R$) or ($M$-$L$) diagram, they are anticorrelated in the CMD.

It is therefore of some value to consider the implications of the age analysis presented
in this section to the comparison of isochrones with the CMD of M4. Figure
\ref{fig:isoCMD} plots the Dartmouth models with the fiducial composition and ages of 10,
11, 12, and 13~Gyr (the same as shown in Figure \ref{fig:iso}). To adjust the isochrones to 
the observed CMD, we adopted a true distance modulus of 11.34 (this is the weighted mean 
from six moduli listed in Table~\ref{abs_parm}), $A_V$=1.47 and $E(B-V)$=0.39. $A_V$ is the
product of E$(B-V)$ and $R_V=3.76$ taken from \citet{hend12}, while $E(B-V)$ itself is taken 
from Kaluzny et al. (in preparation), who derived it for a reference region with uniform 
reddening, in which all stars shown in Figures \ref{fig:cmd} and \ref{fig:isoCMD} 
reside. This derivation is based on two independent sets of observations from 2002 and 2003.
For each season they selected the best photometric night, during which over 50 measurements 
of Landolt standards were made. The agreement with $V$-band magnitudes of M4 stars 
published by \citet[][2012 CADC online edition]{ste00} was excellent, however the
measured  $(B-V)$ was 
on the average  0.023 mag larger than that of Stetson (0.018 mag and 0.028 mag, respectively, 
for the 2002 and 2003 seasons), causing an analogous increase in the derived $E(B-V)$. \citet{hend12} 
obtained a slightly lower reddening of 0.37 mag. While small differences in the 
photometric calibration or in the areas selected for the analysis can easily account for this 
discrepancy, the latter value is inconsistent with the overall agreement shown in 
Figure~\ref{fig:isoCMD}. The isochrone comparison shown in Figure~\ref{fig:isoCMD} is
consistent with an age $\ga 12$ Gyr, higher than derived from the binaries. 
We further discuss the age derived for M4 in Section \ref{sect:discus}.

We define the aggregate age of M4 as an average of the ages of four stars forming V66 and V69 
systems. We calculated two averages. The first is a standard weighted one, including data 
from fits in both the ($M$-$R$) and ($M$-$L$) planes. It is equal to 11.10$\pm$0.26 and 
11.23$\pm$0.27 Gyr, respectively, for DSED and VR isochrones. The second average is obtained 
from ($M$-$R$) fits only, and in two steps. In the first step the mean age of each system is found
by averaging the ages of the components (as explained above, this procedure  removes 
effects resulting from the anticorrelation of the stellar radii). In the second step, a weighted
average of the ages of the two systems is calculated, with the weights equal to inverse errors
squared. For DSED and VR isochrones this procedure 
yields  11.25$\pm$0.42 and 11.30$\pm$0.44 Gyr, respectively, i.e. values entirely compatible 
with those obtained with the first method. 
The final (and conservative) age estimate is given 
by the largest range of ages resulting from the both methods: we may say that M4 is older than 
10.8 Gyr, but younger than 11.7 Gyr, with the most probable age between 11.2 and 11.3 Gyr. 
%We adopt a statistical error of $\pm$0.35 Gyr for the age estimate, the average of these 
%two estimates. 

Following \citet{tho10}, and accounting for the age sensitivities listed 
in Table 14, we adopt a systematic error of 0.85 Gyr arising from a 0.1 dex uncertainty in
each of [Fe/H] and [alpha/Fe]. Based on DSED isochrones, our formal age estimate for M4 derived 
from the study of the binary stars V66 and V69 is 11.25 $\pm$0.42 $\pm$0.85 Gyr. We note 
parenthetically that the He 
abundance of M4 of Y = 0.29 +/-0.01 measured by Villanova et al. (2012) in six blue
HB stars implies an age of approximately 8 Gyr (see Table 14). A second burst of star 
formation occurring after a such a long delay seems very unlikely. 
%Moreover, no evidence
%for any significant age spread is seen in the dereddened CMD of M4 
%published recently by \citet{mil12}. 

\section{Discussion and summary} 
\label{sect:discus}
We have derived absolute parameters of the components of V65, V66 and V69 - three detached eclipsing 
binaries located on the main sequence of the globular cluster M4. The accuracy of
our mass and radii measurements is better than 1.5\% for V65, and better than 1\% for the remaining
two DEBs. The reason for the lower accuracy of the  parameters of V65 is the high activity of this 
system which causes its light curve to be strongly variable (see Figure \ref{fig:V65all}). V65 is a fast 
rotating, X-ray active binary whose components are most probably puffed-up due to the presence of starspots 
and/or magnetic fields, as it is often observed in short-period binaries of spectral type K and M 
\citep[see e.g.][]{tor10a}. These properties, while interesting by themselves, make this object 
unsuitable for analyzes based on isochrone fitting. We note that chromospherically active 
components of eclipsing binaries appear to be larger and cooler than inactive single stars of the same 
mass, but they have a similar luminosity \citep{mor08}. This naturally explains why the primary of V65 is 
(and the secondary may be) located to the red of the main sequence of the cluster.

Based on the parameters of the remaining two systems and two sets of theoretical isochrones obtained 
for Y=0.25, [Fe/H]=$-1.2$ and [$\alpha$/Fe]=$+0.4$, we set lower and upper limit of the age of M4 at 
10.8 and 11.7~Gyr with a formal value of 11.25$\pm$0.42 (statistical) $\pm$0.85 (systematic) Gyr. 
The isochrone comparison shown in Figure~\ref{fig:isoCMD} is
consistent with an age $\sim$12 Gyr.
An age in excess of 12 Gyr has also been derived for M4 by \citet{han04} 
(from fitting of the white dwarf cooling sequence; 12.1 Gyr), \citet{dotter10} (from CMD fitting 
based on ACS data; 12.5 Gyr) and \citet{hend12} (from CMD fitting based on NTT/SOFI data; 12 Gyr 
for [Fe/H]=$-1.0$ -- for a lower metallicity their age would be older). The data listed in Table 
\ref{tab:agsen} suggest that this ``CMD-DEB discrepancy'' might be removed by a slight increase 
in [Fe/H] suggested by the spectroscopic measurements of \citet{mar08} and \citet{vil12}. 
It could also be removed by adopting modest variations in the fiducial helium 
content, [$\alpha$/Fe], or a combination of all three effects. 

We feel, 
however, that the accuracy of the observational data is still too low, and inherent uncertainties 
in the theory of stellar evolution are still too high, to turn such suggestions into firm statements 
concerning the chemical composition of M4. The first of these two factors is illustrated in 
Figs.~\ref{fig:cmd} and \ref{fig:isoCMD} by the scatter of points which define the main sequence 
of the cluster, and the second - by the sensitivity of stellar evolution codes to details of the 
chemical composition \citep[see Table~\ref{tab:agsen}; further discussion of this issue can be 
found in][]{vandenberg12}. Formally, considering the statistical and systematic errors, 
there is no disagreement between the two age estimates.

%However, if more accurate data and better models show that the slight increase 
%in [Fe/H] mentioned above ($\sim$0.1 dex would be sufficient) does not resolve the age 
%discrepancy, then we will have to consider the  mounting evidence of multiple populations 
%in GCs as a solution. Spectroscopy indicates a 
%significant range of light element abundances in M4 \citep[e.g.,][]{iva99,mar08}, 
%which ought to coincide with a range of He content \citep{vil12}. It is entirely 
%possible that the binary systems V66 and V69 are not exactly coeval with the bulk of M4, though
%the age sensitivity summarized in Table 14 suggests that none of them is significantly enhanced 
%in He.

We note here that \citet{dot09} found a similar result for the
DEB V69 in 47 Tuc, where the age derived from the properties of the component stars
is $\sim$1~Gyr younger than the age derived from the location of the main-sequence
turnoff (see their Figure 1).
Milone et al. (2012) have used HST photometry to identify multiple main sequences
in the CMD of 47 Tuc, concluding that approximately 60\% of the cluster population
is in the form of stars with He and N enrichment. If the ground-based photometry
averages out these small color differences on the main sequence, then it is difficult
to explain the discrepancy in ages between the CMD fitting and the age derived from
47~Tuc-V69 as stellar He enrichment since the binary appears younger rather
than older than the mean population, which is already apparently enhanced in He.
In the case of M4, \citet{mar08} and \citet{vil11} have also found evidence for two populations within the
cluster, mainly based on the abundances of Na and CN. However there is no structure on
the main sequence of the CMD that might indicate a He abundance spread. 
It is not clear how the abundance distributions of the two populations might influence
age determination through the properties of stellar tracks calculated with the
different abundances, and we are left with no clear explanation of the measured age 
difference other than assumptions about the chemical composition that
define the fiducial models used in the age measurement.

\begin{figure}[ht!]
\centering
\includegraphics[width=\textwidth,bb= 60 331 548 497,clip]{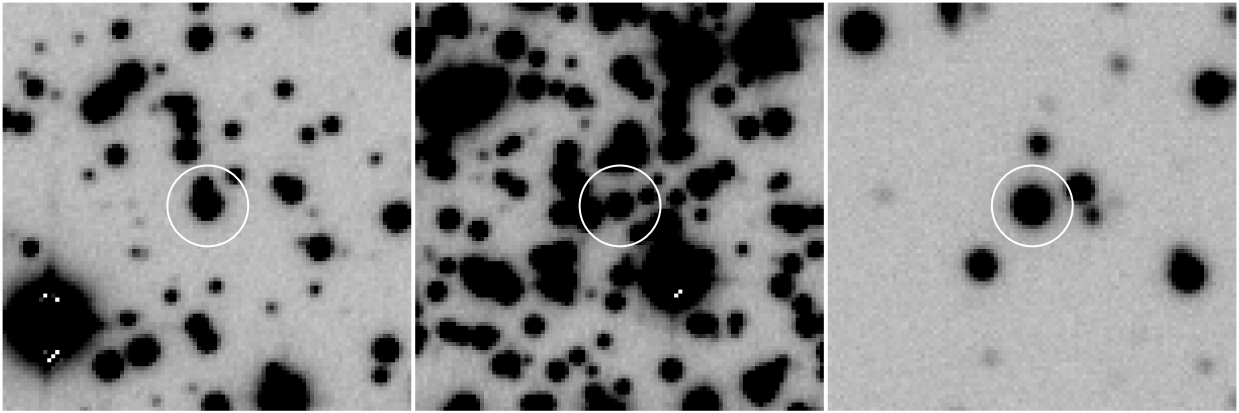}
\caption {Finding charts for V65 (left), V66 (middle) and V69 (right). Each chart
          is 26\arcsec\ on a side, and is oriented with north up and east to the left.
          The coordinates of the targets are given in Table \ref{eqcoord}.
         }
\label{fig:charts}
\end{figure}

\begin{figure}[h]
\centering
\includegraphics[width=0.6\textwidth,bb= 18 342 564 690,clip]{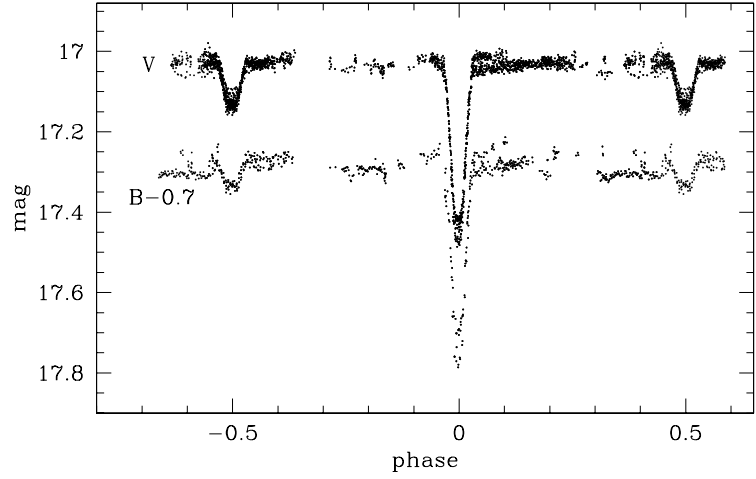}
\caption {Complete photometric observations of V65, illustrating the instability 
          of the light curve. The median internal errors in $B$ and $V$ are, respectively,
          0.009 mag and 0.007 mag. Note that to bring the data closer together 
          on the plot the $B$-curve has been shifted  to  lower values by 0.7 mag.
         }
\label{fig:V65all}
\end{figure}

\begin{figure}[t]
\centering
\includegraphics[width=0.70\textwidth,bb= 41 168 563 690,clip]{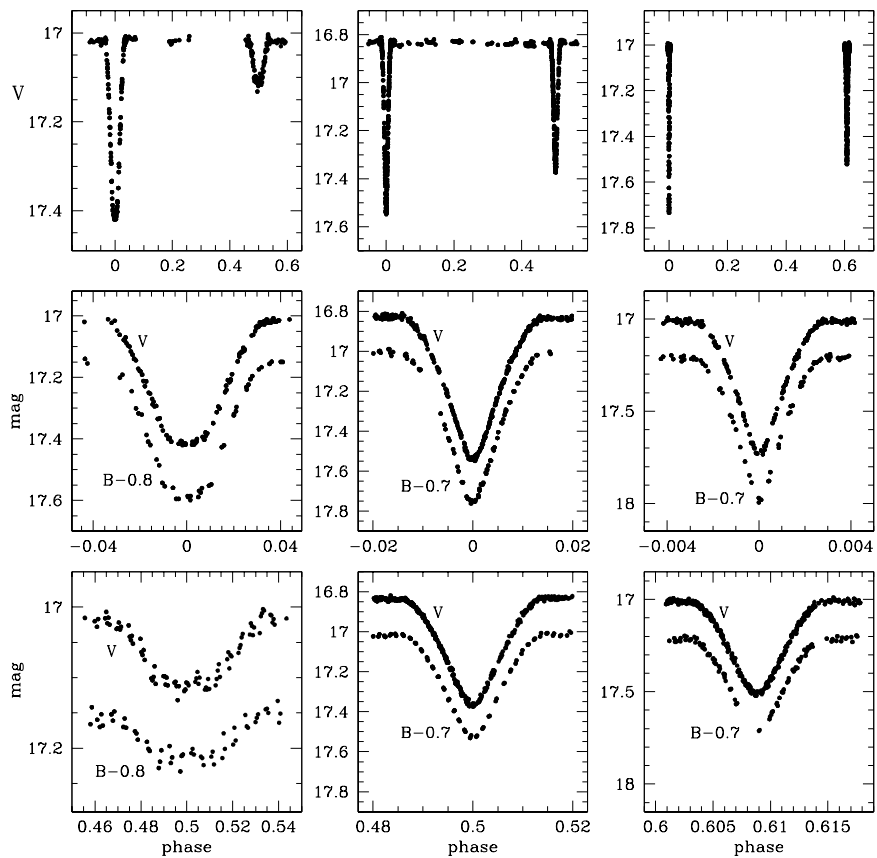}
\caption {Light curves adopted for the analysis: V65 (left), V66 (middle) 
          and V69 (right). The light curve of V65 only includes observations 
          collected between 2008~June~07 and 2009 June 30.  Note that to bring
          the data closer together on the plot the
          $B$-curves have been  shifted to lower values by 0.8 mag (V65)
          and 0.7 mag (V66, V69).
         }
\label{fig:3lc}
\end{figure}

\begin{figure}[t]
\centering
\includegraphics[width=0.30\textwidth,bb= 29 144 564 692,clip]{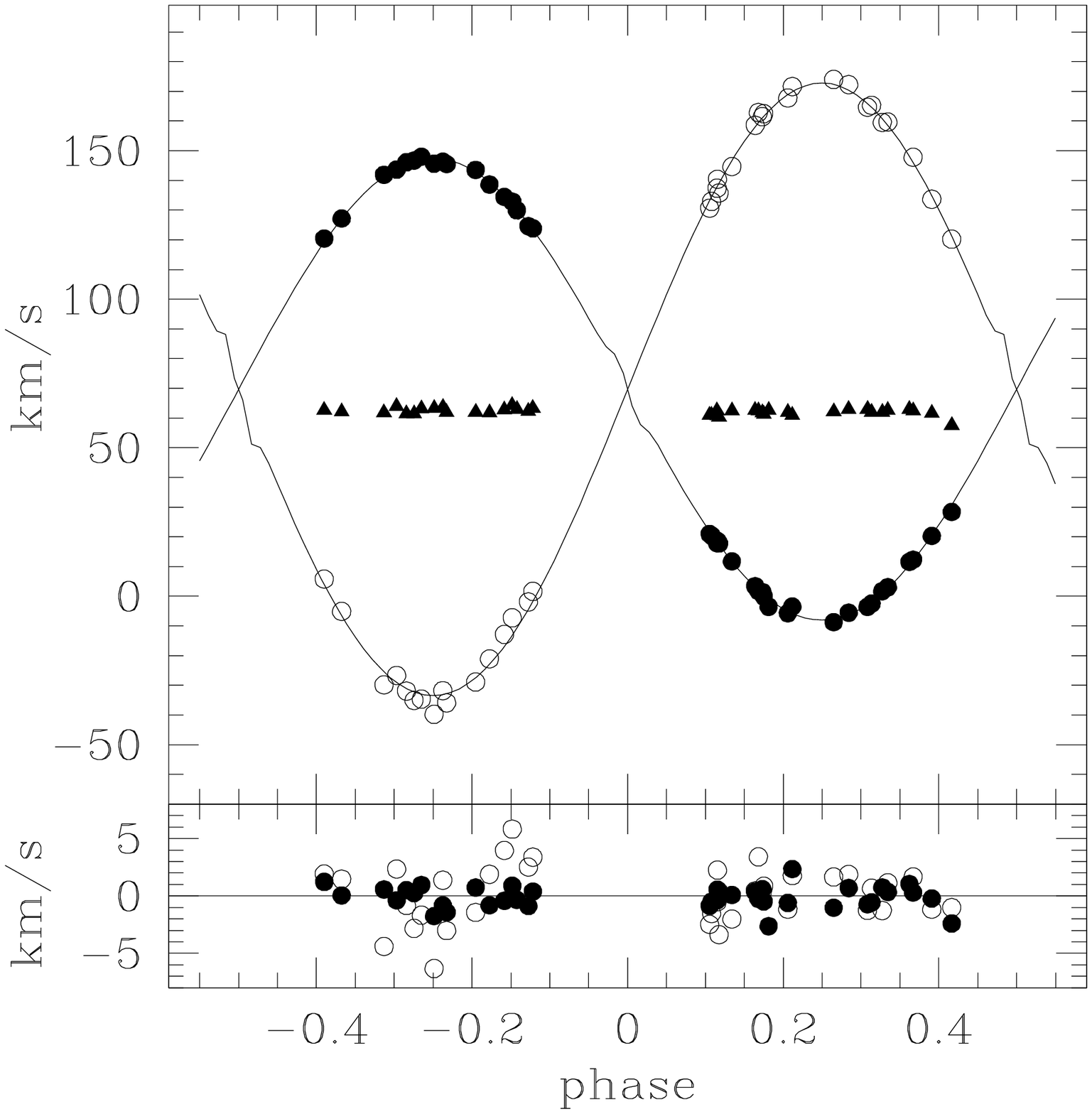}
\includegraphics[width=0.30\textwidth,bb= 29 144 564 692,clip]{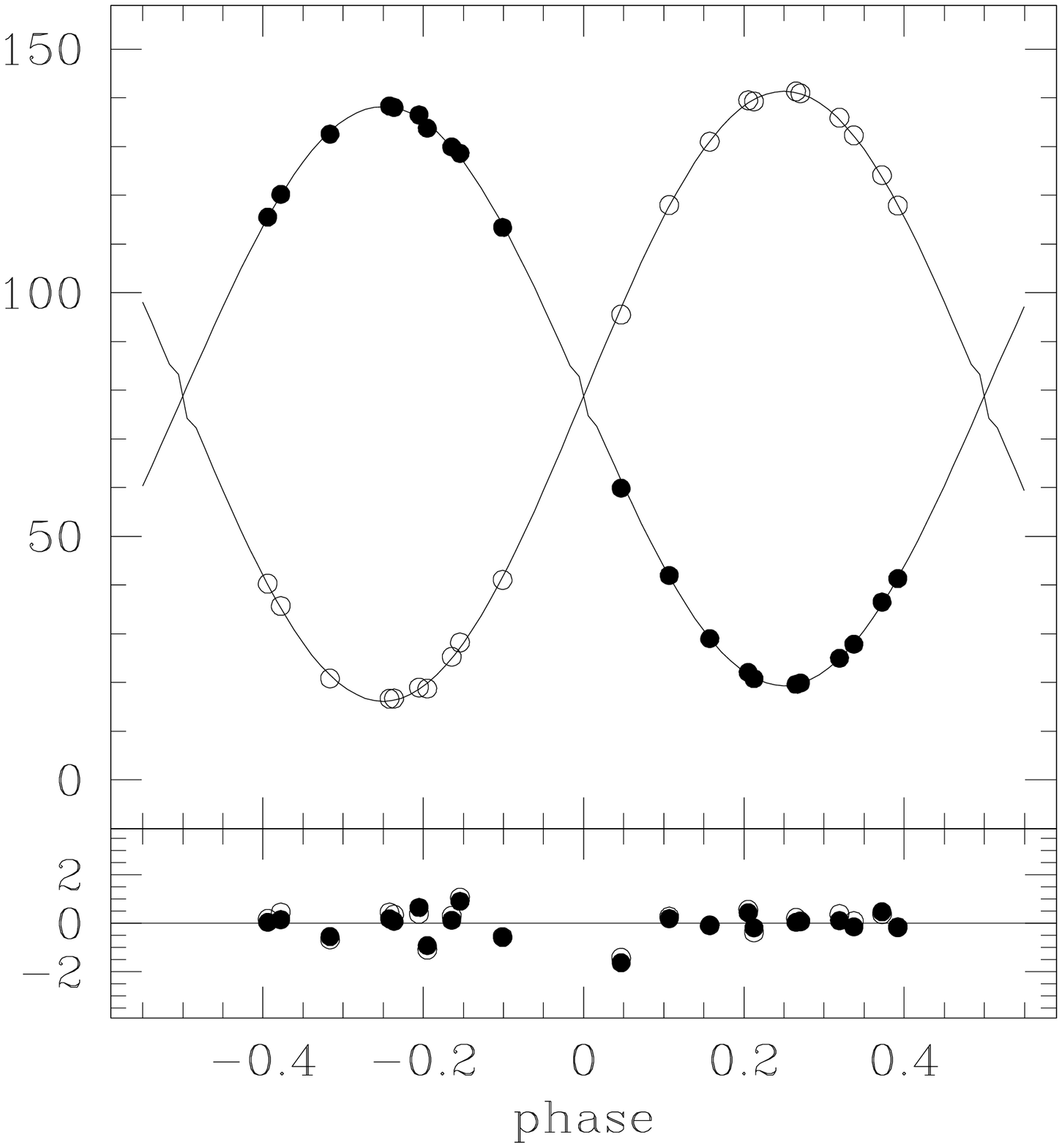}
\includegraphics[width=0.30\textwidth,bb= 29 144 564 692,clip]{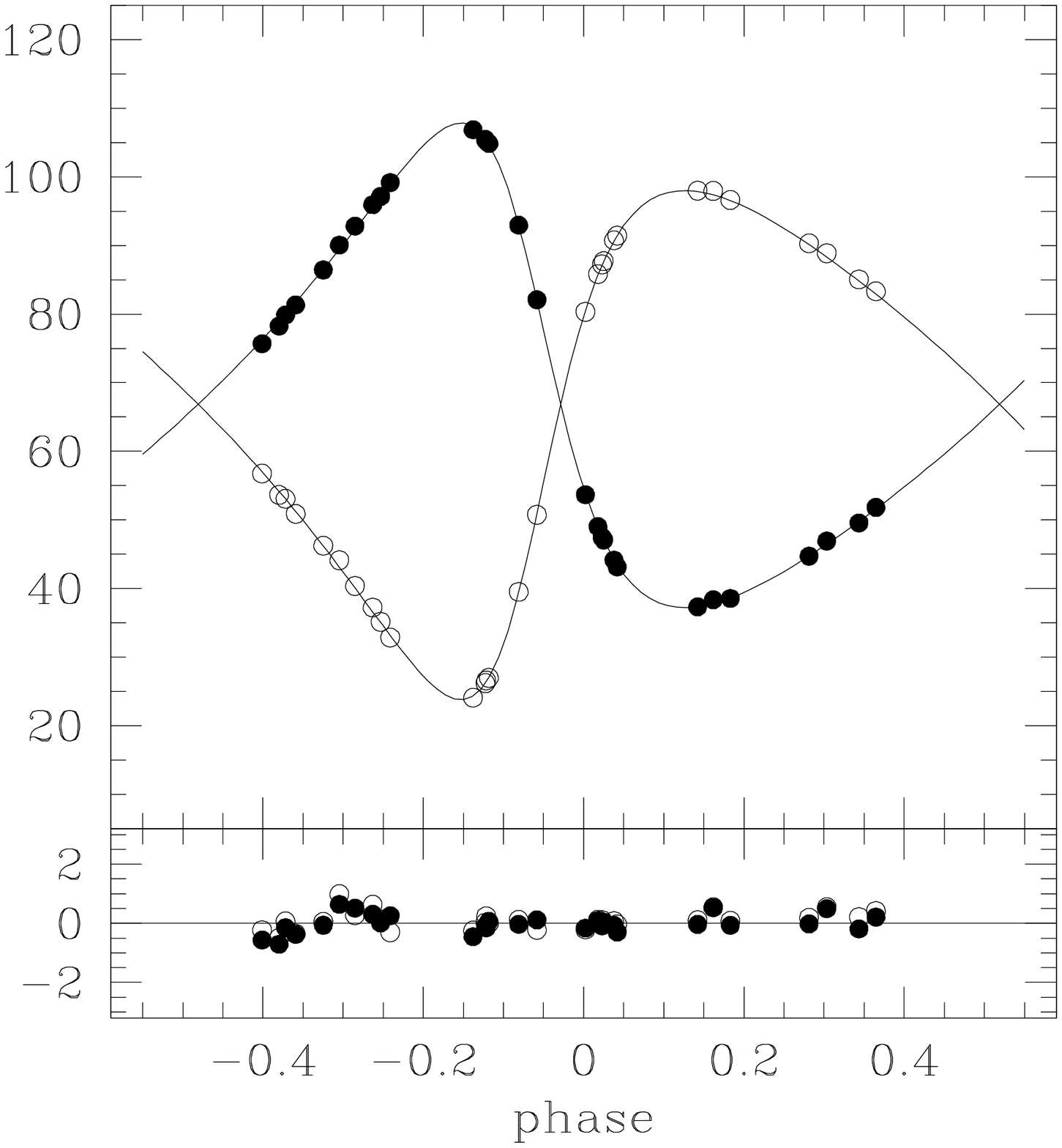}
\caption {Velocity curves adopted for the analysis: V65 (left), V66 (middle) 
          and V69 (right). Since the spectrum of V65 was contaminated by one 
          of the stars blended with the target, the measurements were performed with 
          a three-dimensional extension of the original TODCOR algorithm. The velocities
          of that star are marked in the left panel with triangles. 
         }
\label{fig:3vc}
\end{figure}

\begin{figure}[!t]
\centering
\includegraphics[width=0.90\textwidth,bb= 40 301 564 690,clip]{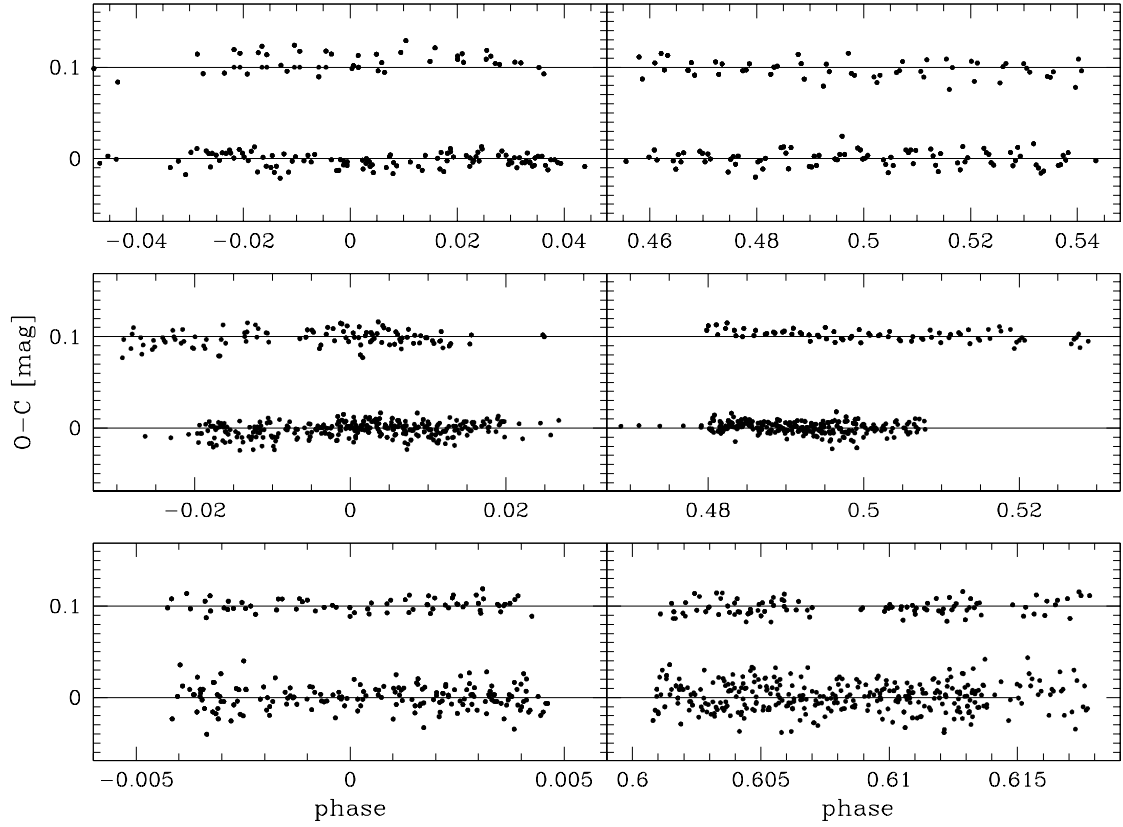}
\caption {Residuals to the fits to the light curves adopted for the analysis: V65 (top), 
          V66 (middle) and V69 (bottom). In each panel, the lower sequence 
          represents the $V$-residuals, and the upper one - the $B$-residuals offset by 
          0.1 mag for clarity.
         }
\label{fig:all_vars_res}
\end{figure}

\begin{figure}[!ht]
\centering
\includegraphics[width=0.32\textwidth,bb= 19 144 564 690,clip]{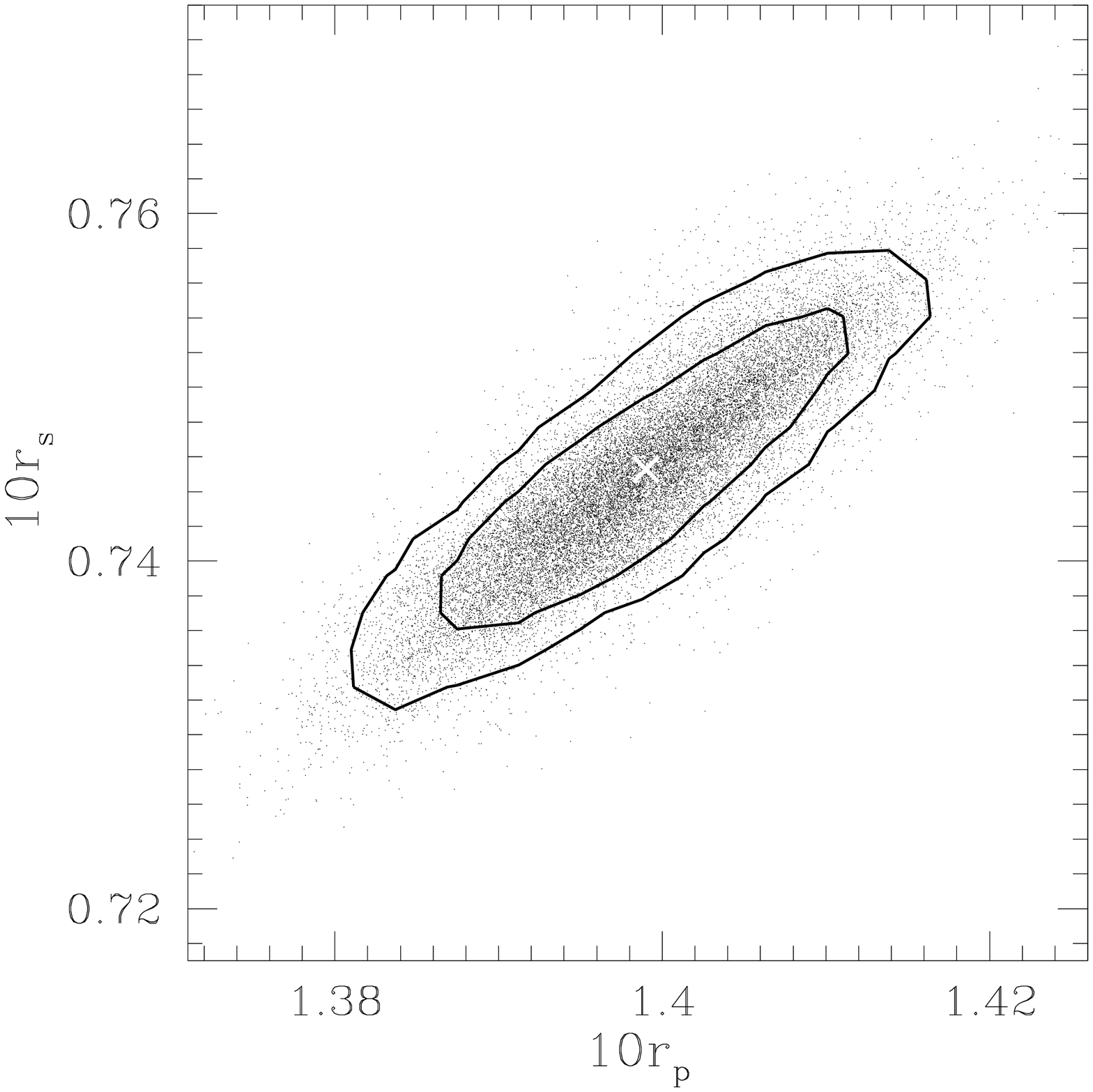}
\includegraphics[width=0.32\textwidth,bb= 19 144 564 690,clip]{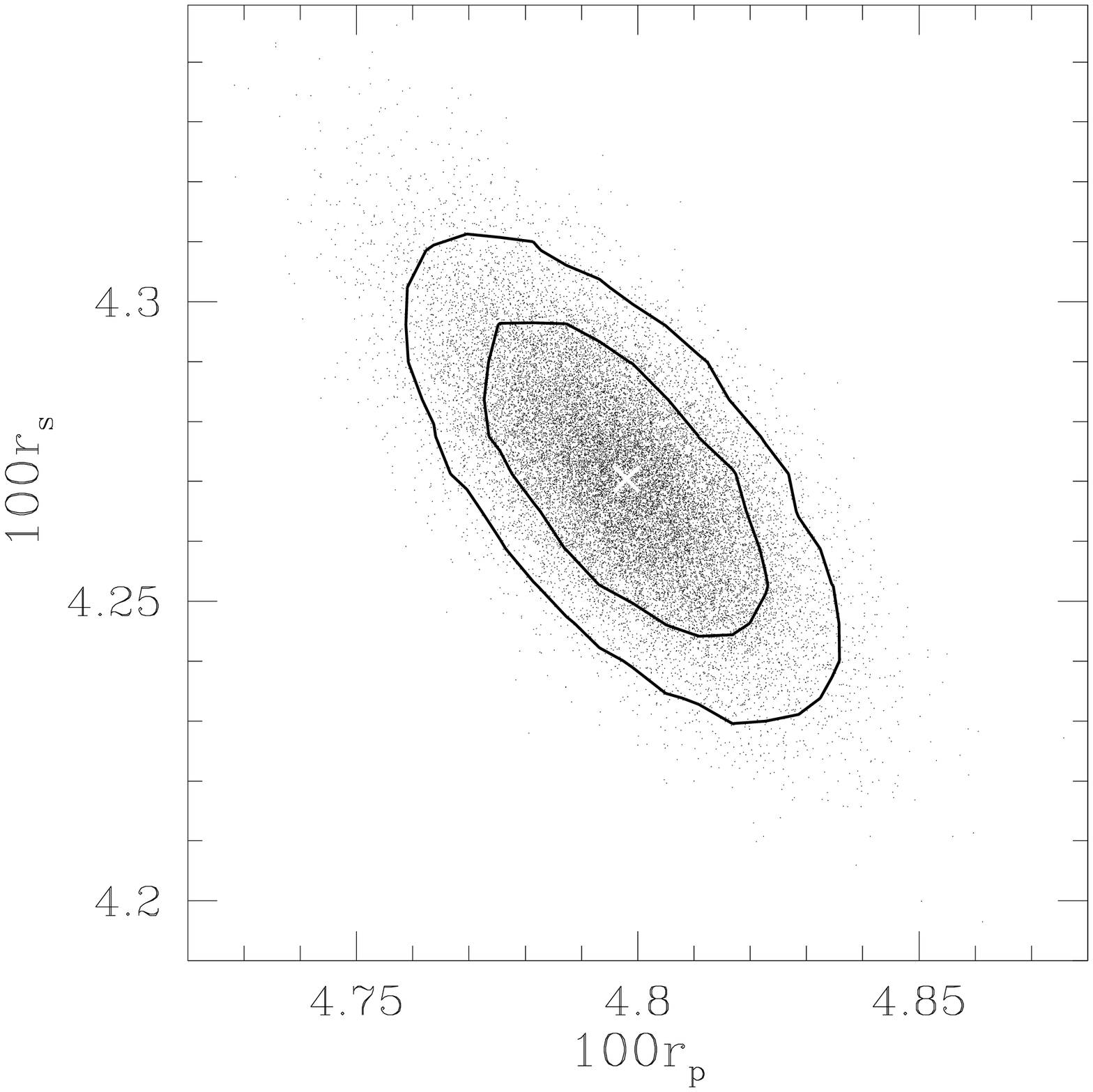}
\includegraphics[width=0.32\textwidth,bb= 19 144 564 690,clip]{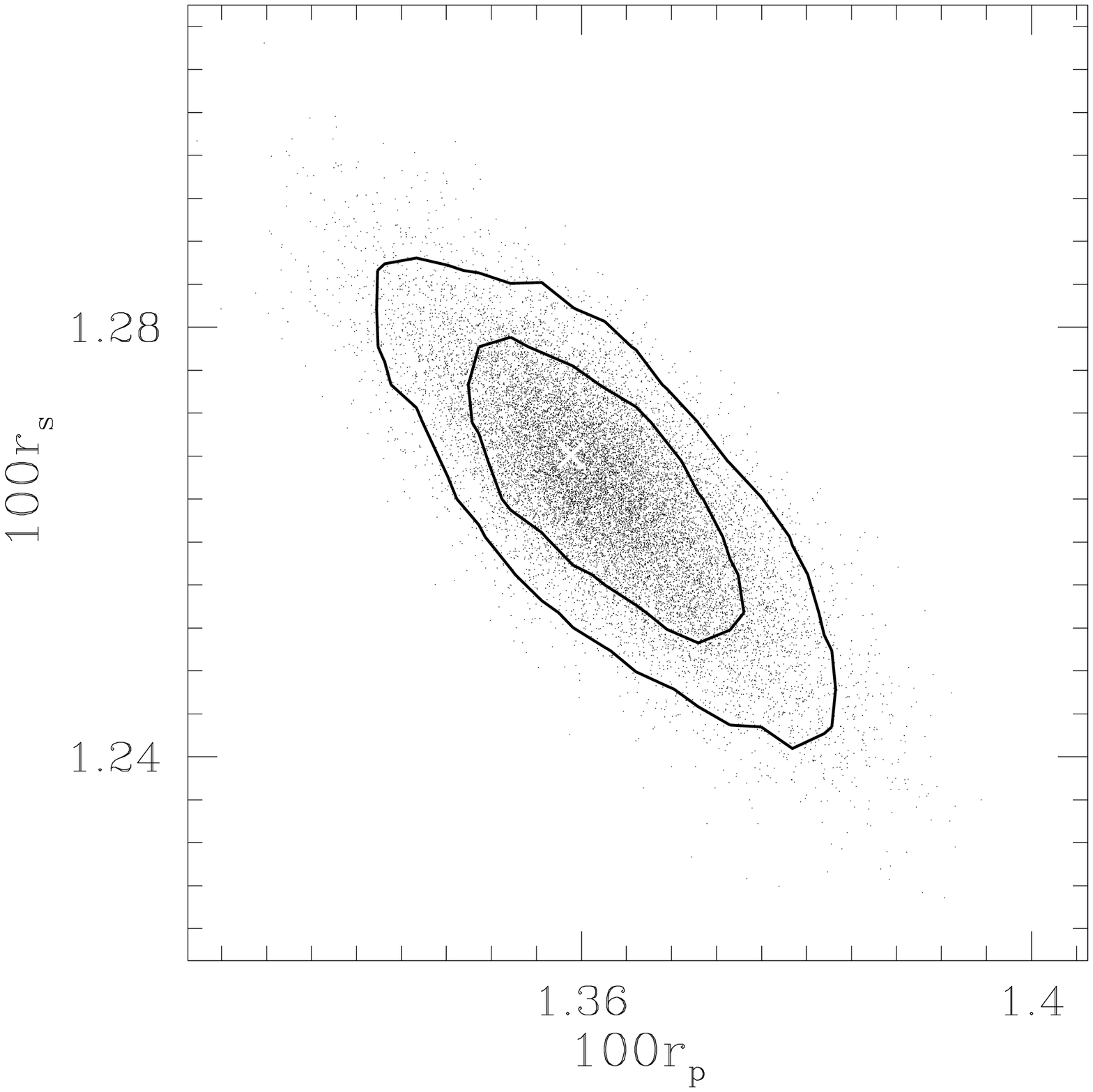}
\caption {Relative radii of the components in V65 (left), V66 (middle) 
          and V69 (right). White crosses: light curve solutions from Table \ref{phot_parm}. 
          Points: results of
          Monte Carlo simulations described in Section \ref{sect:lca_sp}. Inner 
          contour: confidence level 67\%. Outer contour: confidence level 93\%.
          Whenever the errors were not symmetric, the larger one was listed in
          Table \ref{phot_parm}, and used for further error analysis.
         }
\label{fig:3err}
\end{figure}

\begin{figure}[!h]
\centering
\includegraphics[width=0.55\textwidth,bb= 35 145 563 690,clip]{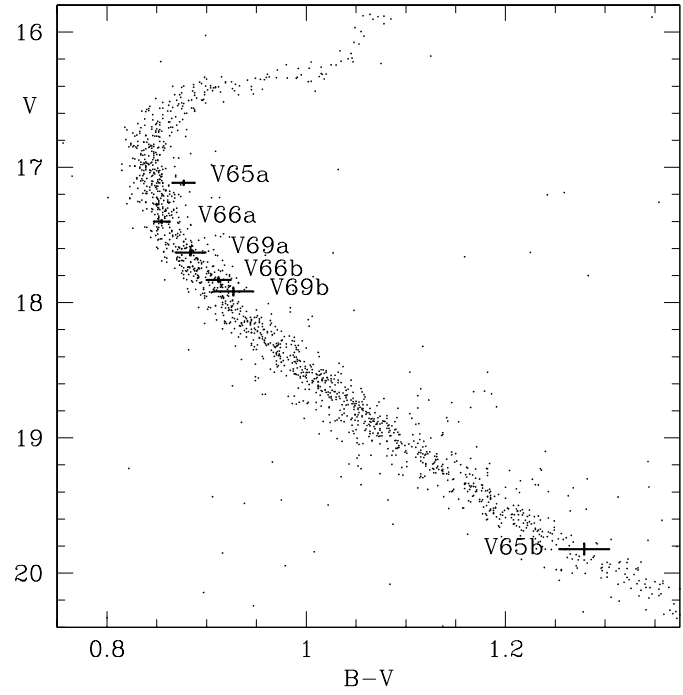}
\caption {The observed color-magnitude diagram of M4 with locations of the components 
          of the three systems investigated in the present paper. Data are taken from 
          Kaluzny et al. (in preparation).
         }
\label{fig:cmd}
\end{figure}

\begin{figure}[!h]
\centering
\includegraphics[width=\textwidth,bb= 40 384 567 689,clip]{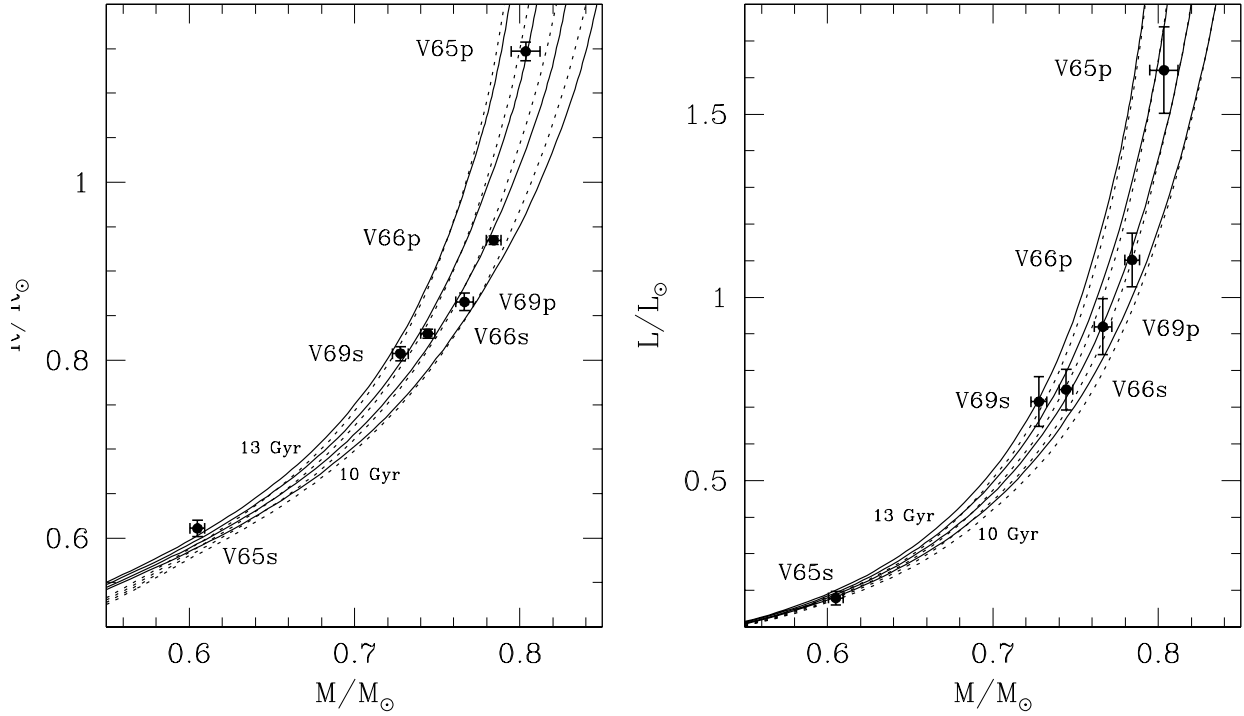}
\caption {Radius-mass and luminosity-mass diagrams with locations of the components 
          of the three systems investigated in the present paper. 
          Solid and dotted lines are Dartmouth and Victoria-Regina isochrones, 
          respectively, for 13, 12, 11 and 10 Gyr from left to right.
         }
\label{fig:iso}
\end{figure}

\begin{figure}[!h]
\centering
\includegraphics[width=\textwidth,bb= 24 260 585 709,clip]{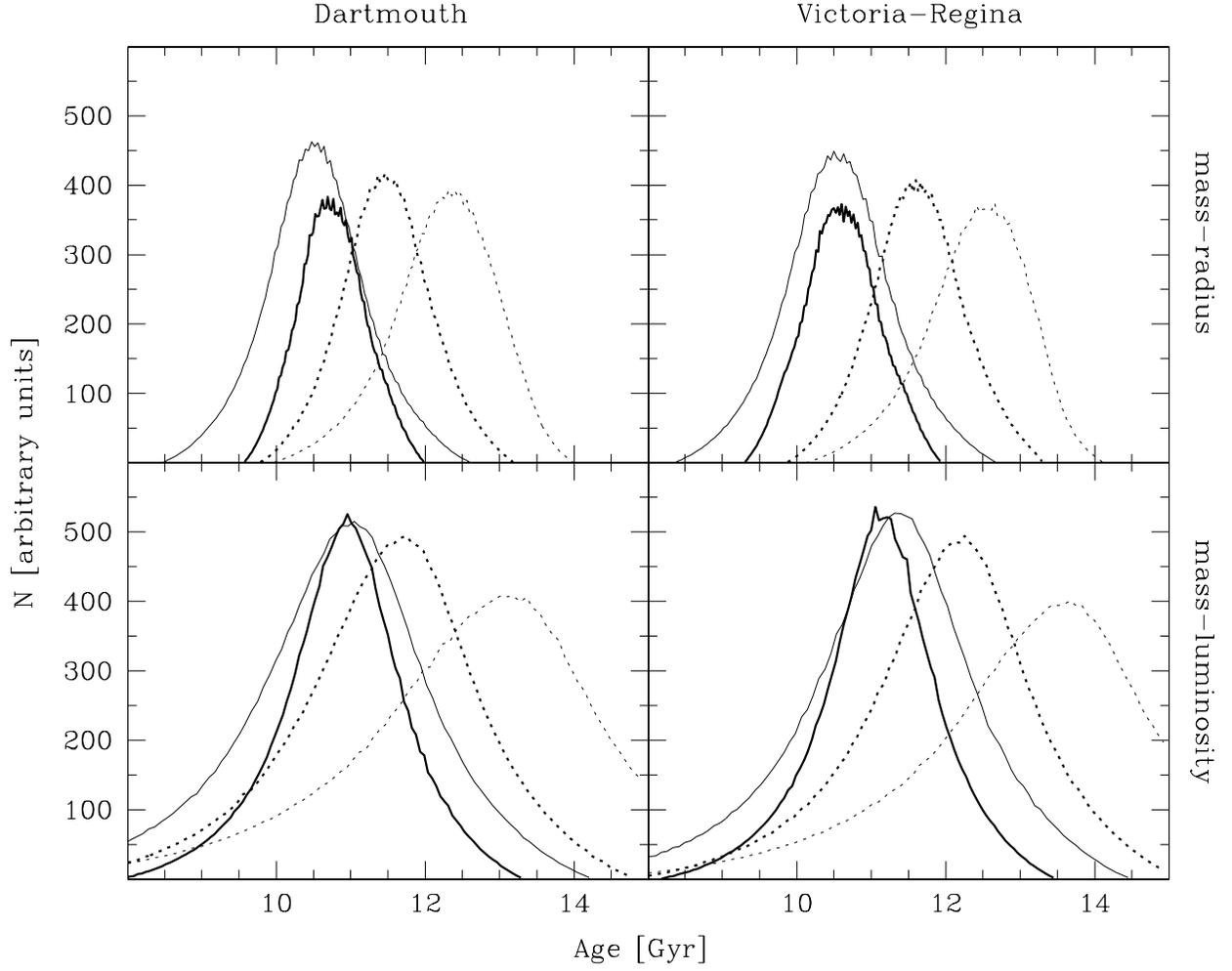}
\caption {Age histograms based on the mass-radius (top) and mass-luminosity (bottom)
analysis for Dartmouth and Victoria-Regina isochrones, both assuming [Fe/H]=$-1.2$,
[$\alpha$/Fe]$\simeq +0.4$, and Y=0.25. Heavy lines: V66 primary (solid) and V66
secondary (dotted). Thin lines: V69 primary (solid) and V69 secondary (dotted).}
\label{fig:hist}
\end{figure}

\begin{figure}[!h]
\centering
\includegraphics[width=0.55\textwidth,bb= 35 145 563 690,clip]{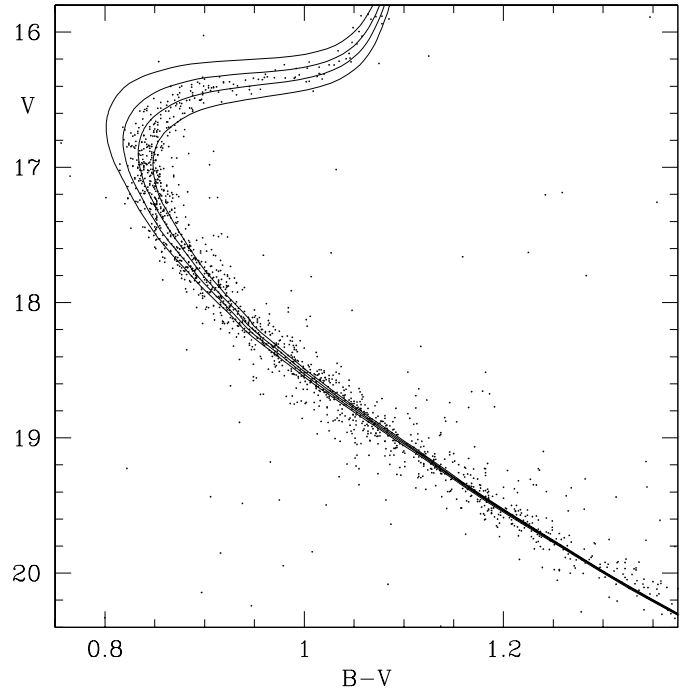}
\caption {The observed CMD of M4 (the same as plotted in Figure \ref{fig:cmd}) compared
with Dartmouth isochrones calculated for the fiducial composition [Fe/H]=$-1.2$,
[$\alpha$/Fe]$\simeq +0.4$, Y=0.25, and ages 10, 11, 12, and 13 Gyr (the same as plotted
in the mass-radius and mass-luminosity planes in Figure \ref{fig:iso}). PHOENIX synthetic 
fluxes were used, and the isochrones were adjusted for a true distance modulus 11.34, 
E$(B-V)$=0.39 and $A_V$=1.47.}
\label{fig:isoCMD}
\end{figure}

How might the accuracy of the observational data be increased?
Given the large inclinations, the errors in the masses of the components of V66 and V69 originate almost
entirely from the orbital solution, which may only be improved by taking additional spectra (preferably 
with the same instrument). This, however, would require a large observational effort, as doubling of 
the present set of radial velocity measurements would lead to an improvement of only 33\% in the mass 
estimates \citep{tho10}.
Contributions to the errors in the radii of the components are dominated by the 
photometric solution, whose accuracy, in turn, depends on the errors of the differential photometry.  
The latter originate mainly from a marginally sampled PSF and limited time-resolution, dictated 
by the diameter of the telescopes we used, and the sensitivity of available detectors. We estimate 
that photometry accurate to 0.002 - 0.003 mag in $V$ would reduce the errors of the radii by 50\%.
Such an improvement is entirely viable, as both V66 and V69 reside in relatively sparsely populated 
areas, and are not blended with another stars (see Figure \ref{fig:charts}). This goal could be 
easily achieved on a 6-8 m class telescope equipped with a camera capable of good PSF
sampling. Better data 
would not remove the anticorrelation of the radii illustrated in Figure~\ref{fig:3err} and briefly
discussed in Section \ref{sect:age} as this is an inherent property of systems with partial eclipses. 
The axes of the error ellipses, however, would become smaller. On the mode\-ling side, detailed 
evolutionary and atmospheric models made specifically to match M4, for which abundant spectroscopic 
information is available, would improve the accuracy of the age analysis.

The luminosities of the components are found using absolute radii and effective temperatures 
estimated from $(B-V)$ - $T_\mathrm{eff}$ calibrations compiled by \citet{cas10}. The errors 
are rather large -- in excess of 0.1$\times\log (L/L_\odot)$. A significant improvement may be expected 
when IR photometry is obtained, and more accurate calibrations linking $(V-K)$ color to surface 
brightness in $V$ are employed.
The anomalous and nonuniform absorption in the field of M4 would still have to be accounted for, 
however the relation between $(V-K)$ and surface brightness is broadly
insensitive to moderate reddening \citep{tho01}. The agreement of the distance modulus derived from 
our DEBS with that recently derived by \citet{hend12} using an 
entirely different method, together with the fit of the observed photometry to model isochrones, 
suggests that these uncertainties are not too high. 

A further observational test would be to determine the chemical abundances of the components of the 
binaries under study using disentangling software \citep[see e.g.][]{had09}. The existing spectra have 
an adequate S/N to measure velocities but not abundances. Given the brightnesses of the components 
and the orbital periods it is possible to obtain higher S/N spectra adequate for abundance analysis
purposes.
\acknowledgments
We thank the anonymous referee for the detailed and helpful report.  
This series of papers is dedicated to the memory of Bohdan Paczy\'nski.
IBT was supported by NSF grant AST-0507325.

\clearpage

%--------------------------------------------------- 

\begin{deluxetable}{lccc}
\tablecolumns{4}
\tablewidth{0pt}
\tabletypesize{\normalsize}
\tablecaption{Times of minima for V65 
   \label{v10989_min}}
\tablehead{
\colhead{E}   &
\colhead{HJD-2450000}     &
\colhead{$\sigma$}    &
\colhead{$O-C$}  
}
\startdata
       1089.0  &  2402.62228 &  0.00024 &  0.00020 \\
       1246.5  &  2763.77804 &  0.00050 & -0.00088 \\
       1397.0  &  3108.88034 &  0.00024 &  0.00020 \\
       1417.0  &  3154.74157 &  0.00019 & -0.00013 \\
       1747.5  &  3912.59451 &  0.00104 & -0.00148 \\
       2058.0  &  4624.58352 &  0.00023 &  0.00018 \\
       2072.0  &  4656.68639 &  0.00027 & -0.00005 \\
       2208.5  &  4969.68786 &  0.00093 & -0.00079 \\
\enddata

\end{deluxetable}

\clearpage

%--------------------------------------------------- 

\begin{deluxetable}{lccc}
\tablecolumns{4}
\tablewidth{0pt}
\tabletypesize{\normalsize}
\tablecaption{Times of minima for V66 
   \label{v11888_min}}
\tablehead{
\colhead{E\tablenotemark{a}}   &
\colhead{HJD-2400000}     &
\colhead{$\sigma$}    &
\colhead{$O-C$}  
}
\startdata
          2.0  &   49916.64227 &        0.00042 &        -0.00001 \\
         39.5  &   50220.81379 &        0.00124 &         0.00234 \\
         40.0  &   50224.87091 &        0.00057 &         0.00088 \\
        129.5  &   50950.83256 &        0.00172 &         0.00089 \\
        174.0  &   51311.78618 &        0.00076 &         0.00027 \\
        174.5  &   51315.84052 &        0.00079 &         0.00158 \\
        218.5  &   51672.73925 &        0.00032 &         0.00020 \\
        219.0  &   51676.79534 &        0.00045 &        -0.00023 \\
        263.0  &   52033.69286 &        0.00029 &        -0.00040 \\
        263.0  &   52033.69260 &        0.00037 &        -0.00014 \\
        308.5  &   52402.75682 &        0.00015 &        -0.00006 \\
        308.5  &   52402.75675 &        0.00025 &         0.00001 \\
        353.0  &   52763.70992 &        0.00023 &        -0.00015 \\
        353.0  &   52763.71045 &        0.00030 &        -0.00068 \\

\enddata
\tablenotetext{a}{Eclipses listed twice were observed in both B and V}

\end{deluxetable}

\clearpage

%--------------------------------------------------- 

\begin{deluxetable}{cccc}
\tablecolumns{4}
\tablewidth{0pt}
\tabletypesize{\normalsize}
\tablecaption{Times of minima for V69 
   \label{v23900_min}}
\tablehead{
\colhead{E}   &
\colhead{HJD-2450000}     &
\colhead{$\sigma$}    &
\colhead{$O-C$}  
}
\startdata
   87   & 4240.72872   &  0.00028  & 0.00044  \\
   93   & 4529.85797   &  0.00032  & 0.00008  \\
101.5\tablenotemark{a} & 4944.69125   &  0.00096  & -0.00042 \\
\enddata
\tablenotetext{a}{The orbit is eccentric, and the secondary minimum occurs
 at phase 0.6086052(77)}
\end{deluxetable}

\clearpage

%--------------------------------------------------- 

\begin{deluxetable}{lccc}
\tablecolumns{4}
\tablewidth{0pt}
\tabletypesize{\normalsize}
\tablecaption{ Equatorial coordinates for three DEBs in M4 (J2000) 
   \label{eqcoord}}
\tablehead{
\colhead{Name}   &
\colhead{RA}     &
\colhead{Dec}    &
\colhead{d\tablenotemark{a}}  \\
\colhead{} &
\colhead{h:m:s} &
\colhead{deg:m:s} &
\colhead{arcmin}
 
}
\startdata
    V65 & 16:23:28.39 & -26:30:22.0 &  1.93 \\
    V66 & 16:23:32.23 & -26:31:41.3 &  0.68 \\
    V69 & 16:23:58.01 & -26:37:18.0 &  7.69 \\
\enddata
\tablenotetext{a}{distance from cluster center at RA = 6:23:35.22,
   Dec = -26:31:32.7. }

\end{deluxetable}

\clearpage

%--------------------------------------------------- 

\begin{deluxetable}{lccccccc}
\tablecolumns{8}
\tablewidth{0pt}
\tabletypesize{\normalsize}
\tablecaption{ HST photometry for V65 (ID = 13372) and two nearby stars }
   \label{and_phot}
\tablehead{
\colhead{Star ID}   &
\colhead{x}     &
\colhead{y}    &
\colhead{$V_{Vega}$}  &
\colhead{err} &
\colhead{$(V-I)_{Vega}$}  &
\colhead{err}  &
\colhead{}
}
\startdata
13372 & 4870.796 & 4386.209 & 16.711 & 0.0022 &  0.942 & 0.0031 & \\
13362 & 4862.341 & 4393.953 & 18.728 & 0.0057 &  1.053 & 0.0078 & \\
13374 & 4877.145 & 4390.062 & 17.490 & 0.0032 &  0.944 & 0.0045 & \\
\enddata

\end{deluxetable}

\clearpage

%--------------------------------------------------- 

\begin{deluxetable}{lccc}
\tablecolumns{4}
\tablewidth{0pt}
\tabletypesize{\normalsize}
\tablecaption{Apparent magnitudes, colors at maximum light, and reddening
   \label{BV}}
\tablehead{
\colhead{Name}   &
\colhead{$V$}     &
\colhead{$B-V$}    &
\colhead{$E(B-V)$}   
}
\startdata
     V65  & 17.028(15) &   0.903(18) & 0.398(10) \\
     V66  & 16.843(12) &   0.878(16) & 0.395(10) \\
     V69  & 17.011(10) &   0.902(10) & 0.403(10) \\
\enddata
\end{deluxetable}

\clearpage

%--------------------------------------------------- 

\begin{deluxetable}{lcccc}
\tablecolumns{5}
\tablewidth{0pt}
\tabletypesize{\normalsize}
\tablecaption{Velocity observations of V65
   \label{v10989vel}}
\tablehead{
\colhead{HJD-2450000 }   &
\colhead{$v_{p}$ [km s$^{-1}$]}     &
\colhead{$v_{s}$ [km s$^{-1}$]}      &
\colhead{$v_{3}$ [km s$^{-1}$]}        &
\colhead{phase} 
}
\startdata
2782.69778 	&	  145.60	&	  -39.80	&	  63.32	&	-0.2487	\\
2783.74040 	&	   -5.70	&	  167.74	&	  62.05	&	0.2060	\\
2868.49631 	&	    1.76	&	  162.88	&	  62.54	&	0.1681	\\
2868.52636 	&	   -3.54	&	  999.99	&	  62.66	&	0.1813	\\
3066.88785 	&	  141.90	&	  -29.85	&	  61.88	&	-0.3131	\\
3067.87522 	&	   17.74	&	  135.74	&	  60.39	&	0.1175	\\
3178.56856 	&	   20.29	&	  133.68	&	  61.69	&	0.3910	\\
3210.52537 	&	    1.73	&	  159.52	&	  61.95	&	0.3274	\\
3517.65000 	&	   -8.72	&	  173.98	&	  62.12	&	0.2649	\\
3517.69441 	&	   -5.55	&	  172.28	&	  62.98	&	0.2842	\\
3518.68412 	&	  146.08	&	  -31.92	&	  61.44	&	-0.2842	\\
3518.72789 	&	  147.96	&	  -34.68	&	  63.16	&	-0.2651	\\
3581.62412 	&	    3.38	&	  158.58	&	  62.62	&	0.1641	\\
3585.55496 	&	  123.85	&	    1.61	&	  63.34	&	-0.1217	\\
3586.60127 	&	    3.04	&	  159.68	&	  62.65	&	0.3346	\\
3587.58254 	&	  146.30	&	  -31.78	&	  63.64	&	-0.2375	\\
3816.80245 	&	  146.68	&	  -35.08	&	  61.54	&	-0.2744	\\
3817.82893 	&	    1.41	&	  161.44	&	  61.80	&	0.1733	\\
3875.71312 	&	   28.39	&	  120.20	&	  57.58	&	0.4166	\\
3877.75800 	&	   -3.58	&	  164.64	&	  63.04	&	0.3084	\\
3891.63975 	&	   11.54	&	  999.99	&	  62.92	&	0.3622	\\
3892.76239 	&	  132.88	&	   -7.28	&	  64.22	&	-0.1482	\\
3937.51145 	&	   12.31	&	  147.79	&	  62.48	&	0.3670	\\
3938.51527 	&	  143.55	&	  -28.89	&	  62.02	&	-0.1953	\\
4139.85710 	&	  120.45	&	    5.80	&	  62.66	&	-0.3899	\\
4259.66316 	&	  129.90	&	  999.99	&	  63.01	&	-0.1423	\\
4314.48934 	&	  145.47	&	  -35.87	&	  61.96	&	-0.2325	\\
4316.63497 	&	  143.63	&	  -26.67	&	  63.99	&	-0.2968	\\
4317.57928 	&	   17.89	&	  137.38	&	  62.48	&	0.1150	\\
4317.62312 	&	   11.73	&	  144.66	&	  62.49	&	0.1341	\\
4328.48847 	&	  124.59	&	   -1.92	&	  62.38	&	-0.1275	\\
4329.49992 	&	   -2.48	&	  165.31	&	  62.00	&	0.3136	\\
4966.64890 	&	   -0.07	&	  162.44	&	  61.40	&	0.1750	\\
4967.69778 	&	  127.16	&	   -5.10	&	  62.14	&	-0.3675	\\
4968.78818 	&	   20.38	&	  132.90	&	  61.06	&	0.1080	\\
5012.59393 	&	   -3.46	&	  171.61	&	  61.02	&	0.2117	\\
5037.57378 	&	   20.97	&	  130.66	&	  61.01	&	0.1055	\\
5354.78703 	&	   37.94	&	  110.54	&	  59.04	&	0.4426	\\
5355.65833 	&	  138.62	&	  -21.08	&	  61.82	&	-0.1775	\\
5355.70217 	&	  134.42	&	  -12.90	&	  62.84	&	-0.1583	\\
5459.51726 	&	   18.60	&	  140.40	&	  60.88	&	0.1155	\\
\enddata
\end{deluxetable}

\clearpage

%--------------------------------------------------- 

%--------------------------------------------------- 

\begin{deluxetable}{lccc}
\tablecolumns{4}
\tablewidth{0pt}
\tabletypesize{\normalsize}
\tablecaption{Velocity observations of V66
   \label{v11888vel}}
\tablehead{
\colhead{HJD-2450000}   &
\colhead{$v_{p}$ [km s$^{-1}]$}    &
\colhead{$v_{s}$ [km s$^{-1}]$}     &
\colhead{phase}
}
\startdata
2736.81000 	&	 132.56	&	   20.83	&	-0.3163	\\
2737.79423 	&	 133.76	&	   18.75	&	-0.1950	\\
2739.75576 	&	  59.89	&	   95.50	&	0.0468	\\
2867.49150 	&	 136.52	&	   18.98	&	-0.2053	\\
3066.84849 	&	  36.52	&	  124.14	&	0.3724	\\
3068.87513 	&	 120.20	&	   35.70	&	-0.3778	\\
3176.56691 	&	 113.41	&	   41.07	&	-0.1010	\\
3178.66358 	&	  29.01	&	  131.01	&	0.1575	\\
3179.53692 	&	  19.63	&	  141.32	&	0.2651	\\
3179.58097 	&	  19.89	&	  140.94	&	0.2706	\\
3180.56655 	&	  41.30	&	  117.88	&	0.3921	\\
3183.53414 	&	 138.30	&	   16.66	&	-0.2421	\\
3183.57940 	&	 138.06	&	   16.72	&	-0.2365	\\
3206.63502 	&	 115.50	&	   40.27	&	-0.3941	\\
3516.81108 	&	 128.63	&	   28.23	&	-0.1541	\\
3520.65104 	&	  24.98	&	  135.92	&	0.3193	\\
3584.61769 	&	  22.07	&	  139.50	&	0.2054	\\
3816.84529 	&	 129.94	&	   25.27	&	-0.1645	\\
3875.82534 	&	  41.97	&	  118.00	&	0.1069	\\
3876.68298 	&	  20.77	&	  139.29	&	0.2126	\\
3877.69367 	&	  27.88	&	  132.30	&	0.3372	\\
\enddata
\end{deluxetable}

\clearpage

%--------------------------------------------------- 

\begin{deluxetable}{lccc}
\tablecolumns{4}
\tablewidth{0pt}
%\tabletypesize{\normalsize}
\tabletypesize{\small}
\tablecaption{Velocity observations of V69
   \label{v23900vel}}
\tablehead{
\colhead{HJD-2450000}   &
\colhead{$v_{p}$ [km s$^{-1}]$}    &
\colhead{$v_{s}$ [km s$^{-1}]$}     &
\colhead{phase}
}
\startdata
3066.80896 	&	  75.68	&	   56.76	&	-0.4011	\\
3067.82945 	&	  78.25	&	   53.64	&	-0.3799	\\
3068.82970 	&	  81.34	&	   50.89	&	-0.3592	\\
3176.61304 	&	 105.44	&	   26.21	&	-0.1224	\\
3176.65755 	&	 105.24	&	   26.59	&	-0.1215	\\
3178.61436 	&	  92.95	&	   39.51	&	-0.0809	\\
3179.71126 	&	  82.09	&	   50.75	&	-0.0581	\\
3182.62446 	&	  53.68	&	   80.34	&	0.0023	\\
3183.62609 	&	  47.47	&	   87.26	&	0.0231	\\
3183.72471 	&	  47.08	&	   87.71	&	0.0251	\\
3184.53725 	&	  43.15	&	   91.44	&	0.0420	\\
3201.52003 	&	  54.00	&	   79.84	&	0.3944	\\
3520.69750 	&	  49.01	&	   85.83	&	0.0180	\\
3521.66112 	&	  44.14	&	   90.75	&	0.0380	\\
3581.57763 	&	  44.71	&	   90.31	&	0.2814	\\
3582.63902 	&	  46.93	&	   88.87	&	0.3034	\\
3584.57357 	&	  49.56	&	   85.06	&	0.3435	\\
3585.59952 	&	  51.81	&	   83.32	&	0.3648	\\
3815.81705 	&	  37.31	&	   97.99	&	0.1423	\\
3816.75737 	&	  38.34	&	   97.94	&	0.1618	\\
3817.78611 	&	  38.55	&	   96.63	&	0.1831	\\
3889.69138 	&	  86.45	&	   46.25	&	-0.3247	\\
3890.65677 	&	  90.06	&	   44.13	&	-0.3046	\\
3891.59731 	&	  92.83	&	   40.37	&	-0.2851	\\
3892.66021 	&	  95.94	&	   37.23	&	-0.2631	\\
3893.71370 	&	  99.17	&	   32.86	&	-0.2412	\\
3898.69475 	&	 106.87	&	   24.08	&	-0.1378	\\
3899.64343 	&	 104.86	&	   26.98	&	-0.1182	\\
3935.60707 	&	  79.87	&	   53.09	&	-0.3718	\\
3989.51380 	&	  97.14	&	   35.16	&	-0.2532	\\
\enddata
\end{deluxetable}
%\clearpage

%--------------------------------------------------- 

\begin{deluxetable}{lccc}
\tablecolumns{4}
\tablewidth{0pt}
\tabletypesize{\normalsize}
\tablecaption{Orbital parameters\tablenotemark{a}
   \label{orb_parm}}
\tablehead{
\colhead{Parameter}   &
\colhead{V65}     &
\colhead{V66}    &
\colhead{V69}   
}
\startdata
     $\gamma$ (km s$^{-1}$)     & 69.55(15)   & 78.76(9)    & 66.90(4)   \\
     $K_p$ (km s$^{-1}$)        & 77.71(19)   & 59.44(14)   & 35.28(9)   \\
     $K_s$ (km s$^{-1}$)        & 103.20(50)  & 62.62(15)   & 37.16(9)   \\
     $e$                       & 0.0\tablenotemark{b} & 0.0\tablenotemark{b} & 0.3840(12)\\
     $\omega$ (deg)            & 0.0\tablenotemark{b} & 0.0\tablenotemark{b} & 65.25(20)   \\
     $\sigma_p$ (km s$^{-1}$)   & 0.99             &   0.56           &  0.33            \\
     $\sigma_s$ (km s$^{-1}$)   & 2.54             &   0.59           &  0.33            \\
     Derived quantities:        &                  &                  &                  \\
     $A\sin i$ (R$_\odot$)     & 8.196(25)  &19.561(35)  &63.681(118)  \\
     $M_p\sin^3 i$ (M$_\odot$) & 0.8024(86) &0.7841(45)  &0.7664(44)   \\
     $M_s\sin^3 i$ (M$_\odot$) & 0.6042(44) &0.7442(42)  &0.7276(43)   \\
\enddata
\tablenotetext{a}{Numbers in parentheses are the errors of the last significant digit(s)}
\tablenotetext{b}{assumed in fit}
\end{deluxetable}

\clearpage

%--------------------------------------------------- 

\begin{deluxetable}{lccc}
\tablecolumns{4}
\tablewidth{0pt}
\tabletypesize{\normalsize}
\tablecaption{Photometric parameters\tablenotemark{a} 
   \label{phot_parm}}
\tablehead{
\colhead{Parameter}   &
\colhead{V65}     &
\colhead{V66}    &
\colhead{V69}   
}
\startdata
     $i$ (deg)     & 88.30(26)   & 89.444(21)    &  89.789(11)    \\
     $r_p$         & 0.1399(12)  & 0.04778(23)   &  0.013591(10)  \\
     $r_s$         & 0.0745(11)  & 0.04242(26)   &  0.012681(13)  \\
     $(L_p/L_s)_V$ & 12.11(28)   & 1.489(20)     &  1.305(42)     \\
     $(L_p/L_s)_B$ & 17.54(50)   & 1.569(22)     &  1.358(44)     \\
     $V_p$ (mag)\tablenotemark{b}   & 17.114(11)(15) & 17.401(8)(13) & 17.629(15)(18) \\
     $V_s$ (mag)\tablenotemark{b}   & 19.822(25)(27) & 17.833(12)(15) & 17.918(20)(22) \\
     $B_p$ (mag)\tablenotemark{b}   & 17.991(11)(15) & 18.256(7)(13) & 18.513(14)(17) \\
     $B_s$ (mag)\tablenotemark{b}   & 21.101(31)(33) & 18.745(11)(15) & 18.845(20)(22) \\
     $\sigma_\mathrm{rms}(V)$ (mmag)
                   & 8             &   7           &  15           \\
     $\sigma_\mathrm{rms}(B)$ (mmag)
                   & 10            &   8           &   8           \\
\enddata
\tablenotetext{a}{Numbers in parentheses are the errors of the last significant digits}
\tablenotetext{b}{For $V_p$, $V_s$, $B_p$ and $B_s$ both the internal error (from the
              photometric solution and profile photometry) and the total error is given,
              the latter including 0.01 mag uncertainty of the zero point of the magnitude 
              scale.}
\end{deluxetable}

\clearpage

%--------------------------------------------------- 

\begin{deluxetable}{lccc}
\tablecolumns{4}
\tablewidth{0pt}
\tabletypesize{\normalsize}
\tablecaption{Absolute parameters\tablenotemark{a}
   \label{abs_parm}}
\tablehead{
\colhead{Parameter}   &
\colhead{V65}     &
\colhead{V66}    &
\colhead{V69}   
}
\startdata
     $A$   (R$_\odot$)  & 8.200(25)  & 19.562(35)  &63.681(118)  \\
     $M_p$ (M$_\odot$)  & 0.8035(86) & 0.7842(45)  &0.7665(53) \\
     $M_s$ (M$_\odot$)  & 0.6050(44) & 0.7443(42)  &0.7278(48)  \\
     $R_p$ (R$_\odot$)  & 1.1470(104) & 0.9347(48) 
%                       & 0.8655$^\mathrm{+0.0097}_\mathrm{-0.0054}$  \\
                        & 0.8655(97)  \\
     $R_s$ (R$_\odot$)  & 0.6110(92) & 0.8298(53) 
%                       & 0.8074$^\mathrm{+0.0066}_\mathrm{-0.0080}$  \\
                        & 0.8074(80)  \\
     $T_p$ (K)          & 6088(108) & 6162(98)  &6084(121) \\
     $T_s$ (K)          & 4812(125) & 5938(105) &5915(137) \\ 
     $L^\mathrm{bol}_p$ 
           (L$_\odot$)  &1.620(118)  &1.129(73) 
%                       &0.920$^\mathrm{+0.076}_\mathrm{-0.074}$        \\
                        &0.920(76)        \\
     $L^\mathrm{bol}_s$ 
           (L$_\odot$)  &0.179(19)  &0.767(55) 
%                       &0.715$^\mathrm{+0.067}_\mathrm{-0.068}$        \\
                        &0.715(68)        \\
     $\log [g_p$ 
           (cm s$^{-2}$)]&4.221(14) & 4.388(77) 
%                       &4.444$^\mathrm{+0.0132}_\mathrm{-0.0093}$ \\
                        &4.444(132) \\
     $\log [g_s$ 
           (cm s$^{-2}$)]&4.645(17) & 4.469(85) 
%                       &4.483$^\mathrm{+0.0105}_\mathrm{-0.0119}$ \\
                        &4.483(119) \\
     $M_{V_p}$ 
           (mag)        &4.329(82)  & 4.716(74)  & 4.945(91)     \\
     $M_{V_s}$ 
           (mag)        &6.983(114) & 5.153(81)  & 5.230(103)      \\
     $(m-M)_{V_p}$ 
           (mag)        &11.400(94) & 11.310(88)  & 11.281(103)     \\
     $(m-M)_{V_s}$ 
           (mag)        &11.454(110) & 11.305(90)  & 11.285(112)      \\

\enddata
\tablenotetext{a}{Numbers in parentheses are the errors of the last significant digits}
\end{deluxetable}

\clearpage

\begin{deluxetable}{ccc}
\tablecolumns{3}
\tablewidth{0pt}
\tabletypesize{\normalsize}
\tablecaption{Results from Isochrone Age Analyses\label{ages}}
\tablehead{\colhead{ID}  & \colhead{Mass-Radius} & \colhead{Mass-Luminosity}\\
           \colhead{}    & \colhead{Age (Gyr)}   & \colhead{Age (Gyr)}}
\startdata
\cutinhead{Dartmouth}
 V65A & $   12.087 \pm    0.712 $ & $   11.618 \pm    0.998 $ \\
 V65B & $   15.068 \pm    1.968 $ & $   10.126 \pm    4.033 $ \\
 V66A & $   10.763 \pm    0.442 $ & $   10.722 \pm    0.861 $ \\
 V66B & $   11.496 \pm    0.587 $ & $   11.217 \pm    1.288 $ \\
 V69A & $   10.570 \pm    0.613 $ & $   10.829 \pm    1.277 $ \\
 V69B & $   12.382 \pm    0.779 $ & $   12.677 \pm    1.701 $ \\
\cutinhead{Victoria-Regina}
 V65A & $   11.797 \pm    0.763 $ & $   11.639 \pm    1.000 $ \\
 V65B & $   16.330 \pm    1.035 $ & $   10.428 \pm    2.104 $ \\
 V66A & $   10.617 \pm    0.491 $ & $   10.948 \pm    0.835 $ \\
 V66B & $   11.638 \pm    0.605 $ & $   11.769 \pm    1.215 $ \\
 V69A & $   10.587 \pm    0.657 $ & $   11.228 \pm    1.215 $ \\
 V69B & $   12.595 \pm    0.800 $ & $   13.224 \pm    1.613 $
\enddata
\tablecomments{These results assume [Fe/H]=$-1.2$, [$\alpha$/Fe]=$+0.4$, and Y=0.25.
               All ages are given as mean $\pm$ standard deviation derived from the
               age histograms presented in Figure \ref{fig:hist} (see text for discussion).}
\end{deluxetable}

\clearpage

%--------------------------------------------------- 
\clearpage

%#For canonical mix, V-R: Z=2.4e-3, DSED: Z=2e-3
%#For scaled-solar mix, both give Z~ 1e-3
%#[Fe/H]=-1.0, [alpha/Fe]=+0.4, Y=0.25 : 0.9 Gyr
%#[Fe/H]=-1.2, [alpha/Fe]= 0.0, Y=0.25 : -1.0 Gyr 
%#[Fe/H]=-1.2, [alpha/Fe]= 0.0, Y=0.25 : -0.4 Gyr
%#[Fe/H]=-1.2, [alpha/Fe]=+0.4, Y=0.27 : -1.7 Gyr
%#[Fe/H]=-1.2, [alpha/Fe]=+0.4, Y=0.29 : -3.2 Gyr

\begin{deluxetable}{ccccc}
\tablecolumns{5}
\tablewidth{0pt}
\tabletypesize{\normalsize}
\tablecaption{Age Sensitivity of the Mass-Radius Plane to Chemical Composition\label{agecomp}
 \label{tab:agsen}}
\tablehead{\colhead{[Fe/H]} & \colhead{[$\alpha$/Fe]} & \colhead{Y} & \colhead{$\Delta$-age} & \colhead{Model} \\
           \colhead{      } & \colhead{             } & \colhead{ } & \colhead{   (Gyr)    } & \colhead{     } }
\startdata
$-1.0$ &    $+0.4$ & 0.25 & $\phm{+}0.9$ & Dartmouth \\
$-1.2$ & $\phm{+}0.0$ & 0.25 & $-0.4$    & Dartmouth \\
$-1.2$ & $\phm{+}0.0$ & 0.25 & $-1.0$    & Victoria-Regina  \\
$-1.2$ &    $+0.4$ & 0.27 & $-1.7$    & Victoria-Regina  \\
$-1.2$ &    $+0.4$ & 0.29 & $-3.2$    & Victoria-Regina
\enddata
\tablecomments{$\Delta$-age is calculated such that a positive value means that the model with
varied composition yields an older age than the model with the fiducial composition
([Fe/H]=$-1.2$,
[$\alpha$/Fe]$\simeq +0.4$, and Y=0.25).}
\end{deluxetable}

\clearpage

%--------------------------------------------------- 

\clearpage

%--------------------------------------------------- 

\clearpage

%--------------------------------------------------- 

\clearpage

%--------------------------------------------------- 

\clearpage

%--------------------------------------------------- 

\clearpage

%--------------------------------------------------- 

\clearpage

%--------------------------------------------------- 

\clearpage

%--------------------------------------------------- 

\clearpage

%--------------------------------------------------- 

\clearpage

%--------------------------------------------------- 

\clearpage

\end{document}